\documentclass{aa}
\usepackage{graphicx}
\usepackage{rotating}
\usepackage{longtable}
\begin{document}

\newcommand{\as}[2]{$#1''\,\hspace{-1.7mm}.\hspace{.1mm}#2$}
\newcommand{\am}[2]{$#1'\,\hspace{-1.7mm}.\hspace{.0mm}#2$}
\def\approxlt{\lower.2em\hbox{$\buildrel < \over \sim$}}
\def\approxgt{\lower.2em\hbox{$\buildrel > \over \sim$}}
\newcommand{\dgr}{\mbox{$^\circ$}}   
\newcommand{\grd}[2]{\mbox{#1\fdg #2}}
\newcommand{\gsim}{\stackrel{>}{_{\sim}}}
\newcommand{\HI}{\mbox{H\,{\sc i}}}
\newcommand{\HIbf}{\mbox{H\hspace{0.155 em}{\footnotesize \bf I}}}
\newcommand{\HIit}{\mbox{H\hspace{0.155 em}{\footnotesize \it I}}}
\newcommand{\HIsl}{\mbox{H\hspace{0.155 em}{\footnotesize \sl I}}}
\newcommand{\HII}{\mbox{H\,{\sc ii}}}
\newcommand{\IHI}{\mbox{${I}_{HI}$}}
\newcommand{\Jykms}{\mbox{Jy~km~s$^{-1}$}}
\newcommand{\kms}{\mbox{km\,s$^{-1}$}}
\newcommand{\kmsMpc}{\mbox{ km\,s$^{-1}$\,Mpc$^{-1}$}}
\def\lir{{\hbox {$L_{IR}$}}}
\def\lco{{\hbox {$L_{CO}$}}}
\def \ls{\hbox{$L_{\odot}$}}
\newcommand{\LB}{\mbox{$L_{B}$}}
\newcommand{\LBnul}{\mbox{$L_{B}^0$}}
\newcommand{\LBsun}{\mbox{$L_{\odot,B}$}}
\newcommand{\lsim}{\stackrel{<}{_{\sim}}}
\newcommand{\LsunK}{\mbox{$L_{\odot, K}$}}
\newcommand{\LsunB}{\mbox{$L_{\odot, B}$}}
\newcommand{\LsunMsun}{\mbox{$L_{\odot}$/${M}_{\odot}$}}
\newcommand{\LK}{\mbox{$L_K$}}
\newcommand{\LKLB}{\mbox{$L_K$/$L_B$}}
\newcommand{\LKLBnul}{\mbox{$L_K$/$L_{B}^0$}}
\newcommand{\LKLsun}{\mbox{$L_{K}$/$L_{\odot,Bol}$}}
\newcommand{\masq}{\mbox{mag~arcsec$^{-2}$}}
\newcommand{\MHI}{\mbox{${M}_{HI}$}}
\newcommand{\MHILB}{\mbox{$M_{HI}/L_B$}}
\newcommand{\MHILBfr}{\mbox{$\frac{{M}_{HI}}{L_{B}}$}}
\newcommand{\MHILK}{\mbox{$M_{HI}/L_K$}}
\newcommand{\KMS}{\mbox{$\frac{km}{s}$}}
\newcommand{\JYKMS}{\mbox{$\frac{Jy km}{s}$}}
\newcommand{\MHILKfr}{\mbox{$\frac{{M}_{HI}}{L_{K}}$}}
\def \ms{\hbox{$M_{\odot}$}}
\newcommand{\Msun}{\mbox{${M}_\odot$}}
\newcommand{\MsunLsun}{\mbox{${M}_{\odot}$/$L_{\odot,Bol}$}}
\newcommand{\MsunLBsun}{\mbox{${M}_{\odot}$/$L_{\odot,B}$}}
\newcommand{\MsunLKsun}{\mbox{${M}_{\odot}$/$L_{\odot,K}$}}
\newcommand{\MT}{\mbox{${M}_{ T}$}}
\newcommand{\MTLBnul}{\mbox{${M}_{T}$/$L_{B}^0$}}
\newcommand{\MTLBsun}{\mbox{${M}_{T}$/$L_{\odot,B}$}}
\newcommand{\nan}{Nan\c{c}ay}
\newcommand{\tis}[2]{$#1^{s}\,\hspace{-1.7mm}.\hspace{.1mm}#2$}
\newcommand{\Vcor}{\mbox{$V_{0}$}}
\newcommand{\vhel}{\mbox{$V_{hel}$}}
\newcommand{\VHI}{\mbox{$V_{HI}$}}
\newcommand{\vrot}{\mbox{$v_{rot}$}}
\def\la{\mathrel{\hbox{\rlap{\hbox{\lower4pt\hbox{$\sim$}}}\hbox{$<$}}}}
\def\ga{\mathrel{\hbox{\rlap{\hbox{\lower4pt\hbox{$\sim$}}}\hbox{$>$}}}}

 \title{Completing H{\Large \bf I} observations of galaxies in the Virgo Cluster}
  
  \author{G. Gavazzi, \inst{1},
	  A. Boselli, \inst{2},
          W. van Driel\inst{3},
      \and
          K. O'Neil\inst{4},
          } 

  \offprints{G. Gavazzi}

  \institute{Universit\'a degli Studi di Milano-Bicocca, Piazza delle scienze 3, 
             20126 Milano, Italy \\
            \email{giuseppe.gavazzi@mib.infn.it}
      \and   
             Laboratoire d'Astrophysique de Marseille, BP8, Traverse du Siphon, 
             F-13376 Marseille, France \\
            \email{alessandro.boselli@oamp.fr}     	     
       \and
             Observatoire de Paris, Section de Meudon, GEPI, CNRS UMR 8111 
             and Universit\'e Paris 7, 5 place Jules Janssen, F-92195 Meudon Cedex, 
             France \\
            \email{wim.vandriel@obspm.fr}
       \and
              NRAO, P.O. Box 2, Green Bank, WV 24944, U.S.A. \\
            \email{koneil@gb.nrao.edu}
             }

\date{\it 12/9/2004}

  \abstract{ {\rm
High sensitivity (rms noise $\sim 0.5$ mJy) 21-cm \HI\ line observations were made of
33 galaxies in the Virgo cluster, using the refurbished Arecibo telescope, 
which resulted in the detection of 12 objects.
These data, combined with the measurements available from the literature, provide the first set of \HI\ data that is complete 
for all 355 late-type (Sa-Im-BCD) galaxies in the Virgo cluster with
$m_p \leq 18.0$ mag.
The Virgo cluster \HI\ mass function (HIMF) that was derived for this optically selected galaxy sample is 
in agreement with the HIMF derived  for the Virgo cluster from the blind HIJASS \HI\ survey
and is inconsistent with the Field HIMF. 
This indicates that both in this rich cluster and in the general field, neutral hydrogen 
is primarily associated with late-type galaxies, with marginal contributions
from early-type galaxies and isolated \HI\ clouds. The inconsistency between the cluster and the field HIMF
derives primarily from the difference in the optical luminosity function of late-type galaxies in the two environments,
combined with the HI deficiency that is known to occur in galaxies in rich clusters.} 
   }

 \authorrunning{Gavazzi et al.}
 \titlerunning{\HI\ observations in the Virgo Cluster}
 
 \maketitle

\section{Introduction}  
Seen from a 25 year perspective, \HI\ observations of galaxies have provided us with some of the most 
powerful diagnostics of the role of 
the environment in regulating the evolution of late-type galaxies in the local Universe. This includes
the definition of
the "\HI\ deficiency" parameter that measures the lack of gas in individual 
cluster galaxies with respect to their ``undisturbed'' counterparts in the field (e.g. Haynes \& Giovanelli 1984,
Solanes et al. 2001 and references therein).
It is well established that spiral galaxies in rich clusters
which have a normal optical morphology have systematically positive \HI\ deficiency 
parameters, i.e. a significantly reduced \HI\ content (e.g.
Haynes et al. 1984). The pattern of \HI\ deficiency found in spiral galaxies that 
are members of rich, X-ray luminous clusters was interpreted 
as due to the dynamical interaction of the galaxy ISM with the hot cluster IGM (e.g. ram-pressure (Gunn \& Gott 1972),
viscous stripping (Nulsen 1982), thermal evaporation (Cowie \& Songaila 1977) or to the tidal 
interaction with nearby companions (Merritt 1983) and/or with the cluster potential well (Byrd \& Valtonen 1990;
Moore et al. 1996)). 
Since the \HI\ deficiency 
parameter indicates if a particular galaxy has already passed through the densest cluster region, 
it is perhaps the most valuable environmental indicator, as it provides a clear signature of
a galaxy's membership of a rich cluster.\\
The Virgo cluster, due to its proximity to us (17 Mpc), 
has received the most attention in \HI\ studies. 
Various works (e.g. Chamaraux et al. 1980;
Helou et al. 1981; Haynes \& Giovanelli 1986; Hoffman et al. 1989, 2003) provided evidence for 
the presence in the Virgo cluster of a mixture of galaxies with extreme \HI\ deficiencies 
and galaxies with normal \HI\ contents. This, in conjunction with distance estimates from 
the Tully-Fisher (1977) relation, provided 
circumstantial evidence for significant infall onto the Virgo cluster (Tully \& Shaya 1984; 
Gavazzi et al. 1999b, 2002). 
In addition to these single-dish studies of  their global \HI\ 
properties, the detailed mapping of Virgo cluster galaxies with radio synthesis telescopes
(e.g., Warmels 1986; Cayatte et al. 1990) 
provided evidence that \HI\ ablation occurs outside-in, producing a spatial truncation of the \HI\ disks
(Cayatte et al. 1994).\\ 
Practically all \HI\ studies of Virgo galaxies were carried out by pointed observations of individual, optically
selected galaxies. Conversely, the first blind \HI\ survey of a $4^{\circ}\times 8^{\circ}$ area of the Virgo cluster
with the 76m Jodrell Bank multibeam instrument (Davies et al. 2004) resulted in the detection of 2 isolated \HI\ clouds,
besides that of 27 previously catalogued galaxies
above the survey's \HI\ mass limit of $5\times 10^7 M\odot$ for a galaxy with a 50 \kms\ linewidth.
A higher sensitivity, full--cluster blind \HI\ survey of the Virgo cluster is planned for 2005-2006 with the ALFA
multibeam system at Arecibo (http://alfa.naic.edu/). 
In preparation for this survey we decided to complete with the present single-beam Arecibo system
the pointed observations of late-type (Sa-Im-BCD) galaxies with $m_p \leq 18.0$ mag in the Virgo cluster area.
Here we report on the results of these observations which, in conjunction with the previously available
\HI\ data-set, enable us to review the properties of galaxies in this cluster as obtained
from optically selected \HI\ observations. \\
The selection of the cluster targets for \HI\ observations
is described in Section 2, the observations and the data reduction are presented 
in Section 3 and the results in Section 4. A discussion of the \HI\ mass function, as derived from
a complete optically selected sample is given 
in Section 5 and the conclusions are presented in Section 6. 
\begin{figure*}
\centering
\includegraphics[width=19.0cm]{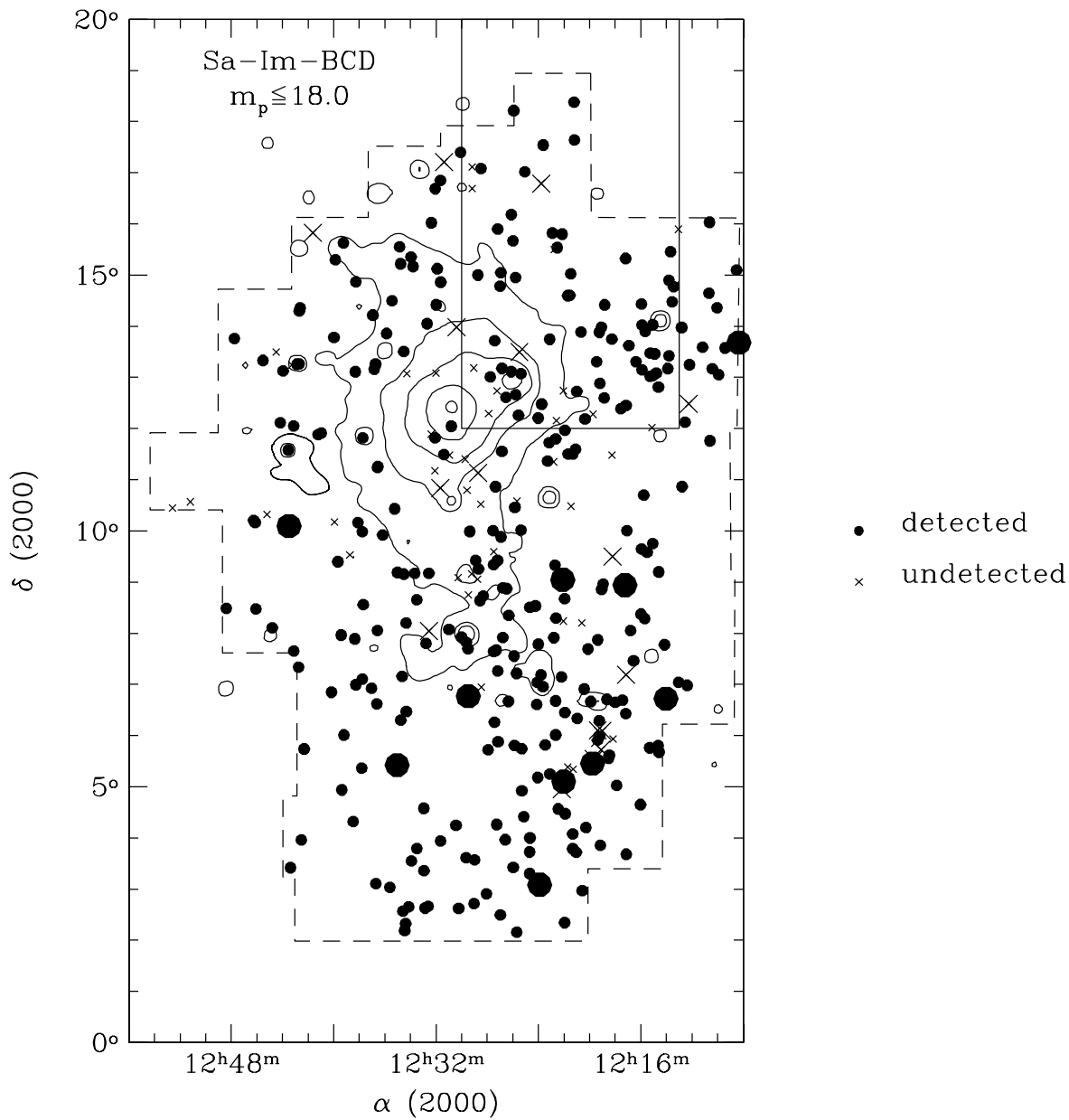}
\caption{The Virgo cluster region considered in the present analysis. The dashed broken line represents
the boundary of the VCC catalog and the rectangle the area covered by the  HIJASS blind HI survey.
Superposed are the X-ray contours from ROSAT (B\"ohringer et al. 1994).
All 355 late-type (Sa-Im-BCD) members of the Virgo cluster with $m_p\leq18.0$ are shown divided into \HI\
detected (296) and undetected (59). Large symbols refer to objects observed in this work.}
\label{sample}
\end{figure*}

\section{Sample selection} 
All data on the Virgo cluster galaxies are collected and made available worldwide 
via the "Goldmine" WWW site (http://Goldmine.mib.infn.it; see Gavazzi et al. 2003). 
The \HI\ completeness of the database is remarkable: the majority ($\sim$ 80 \%)
of disk (Sa-Im-BCD) galaxies with $m_p$ $\leq$ $18.0$ mag belonging to the Virgo cluster
has been detected in \HI\ and for most of the remaining galaxies significant upper limits are available.
The present work is aimed at completing the data on this optically selected sample with 
high sensitivity (rms $\sim 0.5$ mJy) \HI\ observations. 
Our selection criterion includes all 355 disk (Sa-Im-BCD) galaxies with $m_p$ $\leq$ $18.0$ mag that are members,
i.e their measured redshift is in the interval 
$-500<V<3000$ \kms, or bona-fide members of the Virgo cluster, i.e. 26
objects without a direct redshift measurement, that have been classified as belonging to the cluster according to 
the surface brightness criterion used by Binggeli et al. 1985 
in the compilation of the Virgo Cluster Catalog (VCC hereafter), and that were subsequently assigned
to a particular Virgo sub-cloud by Gavazzi et al. (1999b), using a positional criterion.
We include in the target sample all (14) VCC galaxies meeting the above optical
selection criterion that were not observed previously in \HI\, namely:
VCC 1,  99,  227,  256, 275, 315, 517, 528, 531, 675, 679, 1237, 1358 and 1597.
Moreover we targeted the 19 undetected VCC galaxies that were observed previously 
with an rms noise level of 0.7 mJy or higher, namely: VCC 48, 222, 323, 341, 358, 362, 524, 
666, 802, 1086, 1121, 1189, 1196, 1287, 1377, 1435, 1448, 1885 and 1970.\\
A map of the Virgo cluster region is shown in Fig. \ref{sample} where
all (355) galaxies meeting the above selection criterion are shown
with the contours of the X-ray emission from the cluster measured by ROSAT (B\"ohringer et al. 1994) superimposed. 
Including the measurements obtained for this work, all galaxies have been surveyed in \HI\,
with 296 detections and 59 upper limits.\\

\section{Observations} 

Using the refurbished 305-m Arecibo Gregorian radio telescope we observed 57 galaxies
in the Virgo cluster and Coma supercluster (see Section 2) in February 2004, for a total of 28 
hours observing time.
Data were taken with the L-Band Wide receiver, using nine-level sampling with two of 
the 2048 lag subcorrelators set to each polarization channel. All observations were taken 
using the position-switching technique, with the blank sky (or OFF) observation taken for 
the same length of time, and over the same portion of the telescope dish as was used for the 
on-source (ON) observation. Each 5min+5min ON+OFF pair was followed by a 10s ON+OFF 
observation of a well-calibrated noise diode. The overlaps between both sub-correlators 
with the same polarization allowed a contiguous velocity search range while ensuring 
an adequate, wide coverage in velocity. The velocity search range was -1000 to 8500 \kms\. 
The velocity resolution was 2.6 \kms. The instrument's HPBW at 21 cm is 
\am{3}{5}$\times$\am{3}{1} 
and the pointing accuracy is about 15$''$. The pointing positions used
are the optical center positions of the target galaxies listed in Table 1.
Calibration corrections are good to within 10\% (and often much better), see the 
discussion of the errors involved in O'Neil (2004).\\  
Using standard IDL data reduction software available at Arecibo, corrections were applied 
for the variations in the gain and system temperature with zenith angle and azimuth. A 
baseline of order one to three was fitted to the data, excluding those velocity ranges 
with \HI\ line emission or radio frequency interference (RFI). The velocities were corrected 
to the heliocentric system, using the optical convention, and the polarizations were 
averaged. All data were boxcar smoothed to a velocity resolution of 12.9 \kms\ for further 
analysis. For all spectra the rms noise level was determined and for the detected objects 
the central line velocity, the line widths at, respectively, the 50\% and 20\% level of the peak, 
and the integrated line flux were determined.
No flux correction for source to beam size was applied because the optical extent of all detected targets does not
significantly exceed the Arecibo beam.

\section{Results} 

\begin{figure*}
\centering
\includegraphics[width=19.0cm]{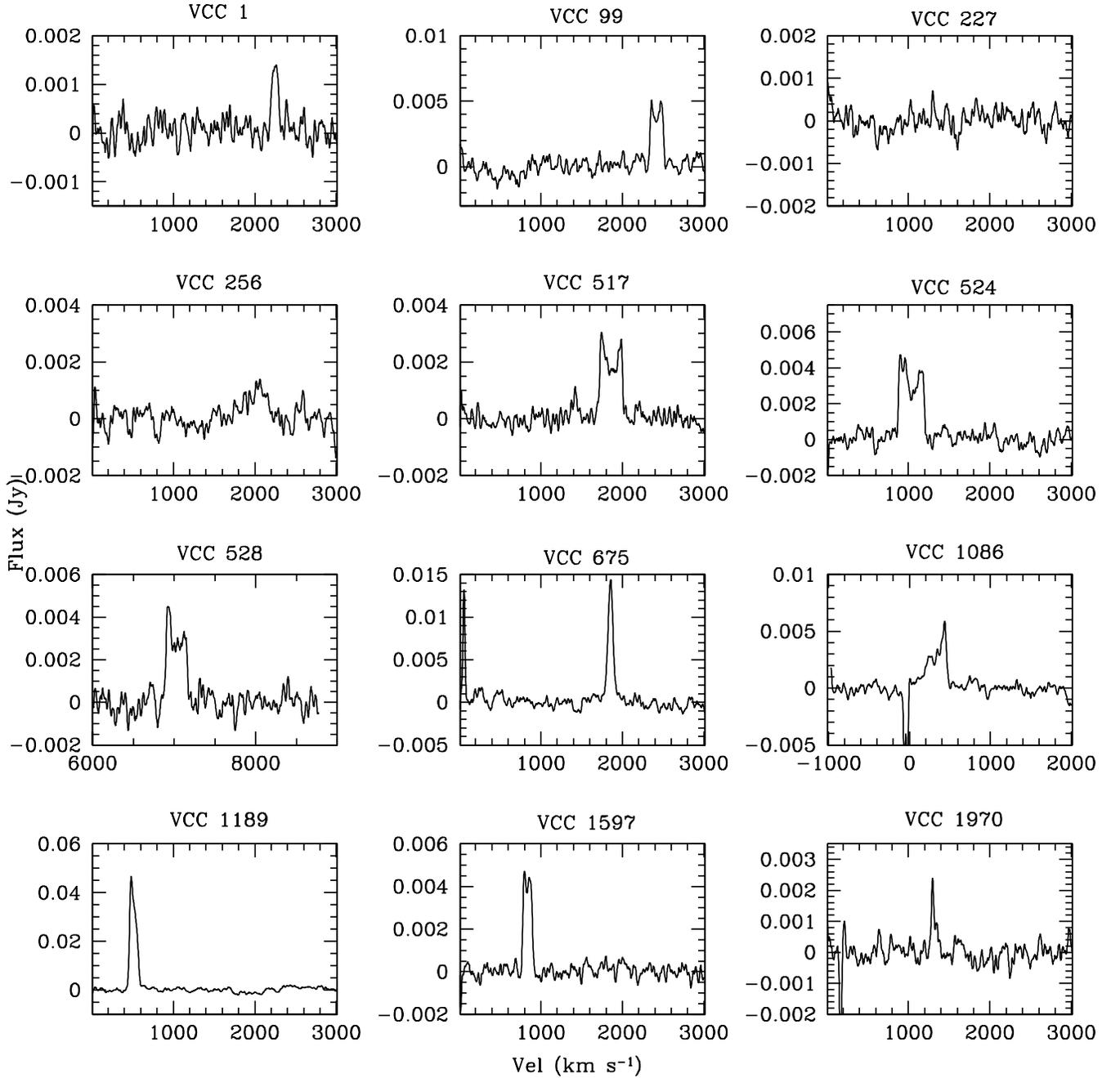}
\caption{\HI\ spectra of the tentatively detected galaxies in the Virgo cluster.}
\label{spectravcc}
\end{figure*}

In order to identify sources whose \HI\ detections could have been confused by nearby galaxies,
we queried the NED, HyperLeda and Goldmine databases and inspected DSS images over a region of 10$'$ 
radius surrounding the central position of each source, given the telescope's sidelobe pattern. 
Quoted values are weighted averages from the HyperLeda database, unless otherwise indicated.\\
The \HI\ spectra of both the clearly and the marginally detected galaxies are shown in 
Figure \ref{spectravcc} 
and the global \HI\ line parameters are listed in Table 1. 
These are directly measured values; 
no corrections have been applied to them for, e.g., instrumental resolution.
Table 1 is organized as follows:\\
Column 1: Obj. is the galaxy designation; \\ 
Column 2-3: (J2000) celestial coordinates;\\
Column 4: the heliocentric optical recessional velocity (in \kms);\\ 
Column 5: the rms dispersion in the baseline (mJy); \\ 
Column 6: $S_p$ is the peak flux of the detected line (mJy); \\ 
Column 7: $V_{HI}$ is the heliocentric central radial velocity of a line profile 
(in \kms), in the optical convention, with its estimated uncertainty (see below); \\
Columns 8-9: $W_{50}$ and $W_{20}$ are the line widths at 50\% and 20\% of peak maximum, respectively, (\kms); \\ 
Column 10: $I_{HI}$ is the integrated line flux (\Jykms), with its estimated uncertainty (see below).\\  
Column 11: A quality flag to the spectra is given, where
Q=1 stands for high signal-to-noise, double horned profiles, 
Q=2 for high signal-to-noise, single horned profiles, and 
Q=3,4 for low signal-to-noise profiles whose measured line parameters are not
reliable.\\ 
We estimated the uncertainties 
$\sigma_{V_{HI}}$ (\kms) in \VHI\ and
$\sigma_{I_{HI}}$ (\Jykms) in \IHI\ following Schneider et al. (1986, 1990), as:
\begin{equation}
\sigma_{V_{HI}} = 1.5(W_{20}-W_{50})X^{-1}
\end{equation}
and 
\begin{equation}
\sigma_{I_{HI}} = 2(1.2W_{20}/R)^{0.5}R\sigma = 7.9(W_{20})^{0.5}\sigma 
\end{equation}
where \IHI\ is the integrated line flux (Jy~\kms), 
$R$ is the instrumental resolution (12.9 \kms), and
$X$ is the signal-to-noise ratio of a spectrum, i.e. the ratio of the 
peak flux density $S_p$ and $\sigma$, the rms dispersion in the baseline (Jy). \\
The uncertainty in the $W_{20}$ and $W_{50}$ line widths is expected to be
2 and 3 times $\sigma_{V_{HI}}$, respectively.\\
Of the 33 observed Virgo cluster objects, 12 (36\%) were detected, of which 4 tentatively (see Table 1). 

\section{Discussion}

The newly obtained \HI\ data were combined with those available from the
literature for the $m_p \leq 18.0$ late-type (Sa-Im-BCD)\footnote{Galaxies of type S..., dE/Im and ``?'' are not
included in the present analysis.} galaxies in the Virgo cluster, listed in Table \ref{sample_dat}.
The sample comprises 355 galaxies, of which 296 were detected and 59 remain undetected.\\
The compilation of \HI\ data from the literature was carried out using a criterion of
maximum reliability and homogeneity, i.e. recent Arecibo data were preferred to more ancient measurements.
In most cases we were able to assess a quality flag to the
quoted measurement by inspecting the individual HI profiles.
It is known that the \HI\ distribution in normal late-type galaxies spatially exceeds 
the optical extent by a factor ranging from 1.2 (Hewitt et al. 1983) to 2 (Salpeter \& Hoffman, 1996) on average.
Not accounting for such an effect makes the measurement of the total \HI\ flux significantly underestimated 
from single pointing observations of
galaxies comparable in size to the beam (see Sullivan et al. 1981, Hewitt et al. 1983).
Although this is not a big concern for galaxies observed in this work, because they are generally 
small compared to the Arecibo beam, it might be a problem for 107/355  
galaxies with optical diameters $>$ 2 arcmin, whose \HI\ parameters have been taken from the
literature. In order to minimize the missing flux problem we took the published HI parameters 
by selecting the references with the following priority:
interferometer or mapping surveys were preferred to single beam pointings.
Among the latter works we first selected those which include the flux correction 
for source to beam size. Only a small fraction of the large galaxies (14/107) 
were taken from references not including such a correction, that was not either applied by us.
The reader should be aware that a possible overestimate 
of the \HI\ deficiency parameter among some of the largest (most luminous) galaxies might arise from this effect.\\
Distances were estimated as in Gavazzi et al. (1999b):
individual objects were assigned to 
the various subclouds in the Virgo cluster according to a positional/velocity criterion:
A=cluster A (M87), B= cluster B (M49), 
W=west cloud, M=M cloud, N=North cloud, E=East cloud, S=Southern extension.
For each cloud a mean distance D is assumed:
17 Mpc for clouds A, E, S and N, 23 Mpc for cloud B, and 32 Mpc for clouds W and M.
We estimate that the distances of individual objects are subject to $\sim$ 30\% uncertainties.\\
The corrected \HI\ flux was transformed into \HI\ mass or mass limit, in solar units, adopting
\MHI = 2.36 $10^5 D^2$ \IHI.
For undetected galaxies we set \IHI=$1.5 \times rms_{HI} \times W_{<20-50>}$,
where rms$_{HI}$ 
is the rms of the spectra in mJy and the W$_{<20-50>}$ profile width is based on the 
following average line widths of the detected objects per Hubble type bin: 
300 \kms\ for Sa-Sbc, 190 \kms\ for Sc-Scd, and 85 \kms\ for Sm-BCD;\\
We also estimate the \HI\ deficiency parameter following Haynes \& Giovanelli (1984)
as the logarithmic difference between $M_{HI}$ of a reference sample of isolated 
galaxies and $M_{HI}$ actually observed in individual objects: $Def_{HI}= Log M_{HI~ref.} - Log M_{HI~obs.}$. 
$Log M_{HI~ref}$
has been found linearly related to the galaxies linear diameter $d$ as: 
$Log M_{HI~ref}=a+b Log(d)$,
where $a$ and $b$ are weak functions of the Hubble type, as listed in Table 3. 
The problem here is that Haynes \& Giovanelli (1984) have included in their
reference sample of isolated galaxies only relatively large ($a>1$ arcmin) UGC objects 
so that the $Def_{HI}$ parameter is poorly calibrated for smaller objects,
making determinations of the HI deficiency for the smallest objects uncertain, 
likely underestimated (Solanes 1996). 
Furthermore, as discussed in Solanes et al. (2001) galaxies in the latest
Hubble types (Scd-Im-BCD), for which we have adopted $a$ and $b$ parameters
consistent with those of Sc (Table 3), are more subject to observational biases
than higher surface brightness galaxies. The reader should be aware that
the determinations of the HI deficiency for these objects is highly uncertain.

\begin{table}
\begin{tabular}{ccc}
\multicolumn{3}{l}{\footnotesize {\bf Table 3} Adopted parameters of the \MHI\ vs. diameter relation.} \\
\smallskip \\
\hline
Type & a & b  \\ 
\hline
Sa-Sab     & 7.17 & 1.64\\ 
Sb         & 7.29 & 1.66\\ 
Sbc        & 7.27 & 1.70\\ 
Sc         & 6.91 & 1.90\\ 
Scd-Im-BCD & 7.00 & 1.88\\
   \noalign{\smallskip}
   \hline
   \end{tabular}
   \setcounter{table}{3}
   \label{deftab}
   \end{table}

\subsection{Comparison with Leda} 

\begin{figure}
\centering
\includegraphics[width=9.0cm]{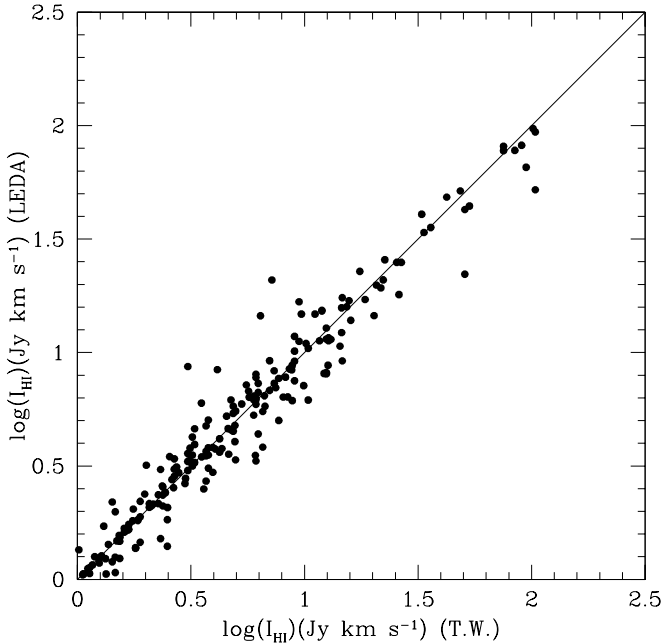}
\caption{
Comparison of the HI fluxes adopted in this work and those found in the Leda database.}
\label{Leda}
\end{figure}

The \HI\ fluxes listed in Table \ref{sample_dat} were compared to the 
\HI\ fluxes listed in the HyperLeda database (http://foehn.univ-lyon1.fr/hypercat/), found for 286 detected galaxies.
The comparison of the two datasets is given in Fig. \ref{Leda}, showing excellent agreement:
$I_{HI}(T.W.) = I_{HI}(Leda)*1.04 \pm 0.25$
The most discrepant objects are VCC 66, 1987 and 2070 for which our flux is almost
twice the flux in Leda, and VCC 1673 showing the reverse ratio.
Assuming that the two data-sets are independent (which is not true because several measurements are in common)
and assuming that the error is equally distributed among the two databases, we estimate that 
the flux uncertainty given in this work is $\sim$ 20 \%. 
Combining this error with the distance uncertainty ($\sim$ 30\%) discussed in the previous Section
we conclude that the uncertainty on the \MHI ~estimates is $\sim$ 50 \%, or $\sim$ 0.2 on log(\MHI). 

\subsection{Comparison with HIJASS}
\label{daviescomp}

\begin{figure}
\centering
\includegraphics[width=9.0cm]{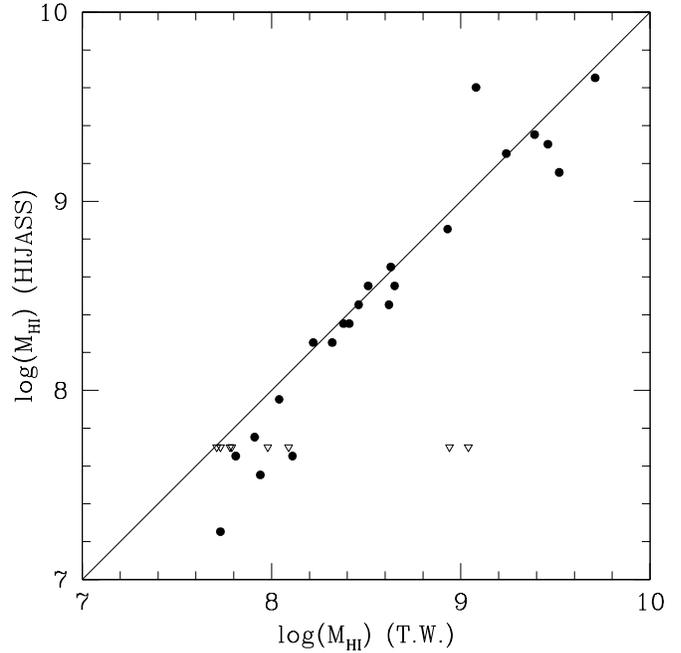}
\caption{
Comparison of the HI masses as derived from this work with those derived 
in the blind HIJASS \HI\ survey (Davies et al. 2004)
for galaxies in common in the radial velocity interval of 500-2500 \kms. 
Open triangles are for undetected HIJASS galaxies.}
\label{hipass}
\end{figure}

It is worth comparing the \HI\ masses listed in Table \ref{sample_dat} with those
obtained by the HIJASS blind Virgo cluster \HI\ survey (Davies et al. 2004, hereafter D04) for the common galaxies 
in the area: $12^h13^m<R.A. (J2000) <12^h30^m; +12^o00'<dec<+20^o00'$.
They are marked as "D04" in the references to Table \ref{sample_dat}.
Out of the 27 galaxies detected by HIJASS only 22 are included in our Table \ref{sample_dat} because 3
are not in the VCC and 2 others, which were considered as spirals by HIJASS, are classified as S0 and "?"
in the VCC, and were thus not considered by us. The remaining 22 galaxies were detected also in the present survey.
Moreover D04 narrowed their search velocity range to 500-2500 \kms ~in order
to avoid Galactic emission and background galaxies. Galaxies 
in common with D04, in the interval 500-2500 \kms, are marked with an asterisk in Table \ref{sample_dat}. 
With these restrictions,  
the HI mass measured by D04, re-scaled to the distance adopted by us (i.e. 17 Mpc for cluster A,
and 32 Mpc for cloud M, instead of 16 Mpc adopted by D04) is plotted vs. 
our mass estimates in Fig. \ref{hipass}, including
galaxies undetected by D04 that are plotted at $M_{HI}=10^{7.7}$ M$\odot$ (triangles). 
Two galaxies 
(VCC 119 = UGC 7249, VCC 483 = NGC 4298) are 
undetected by D04 in spite of their mass in excess of 10$^{8.9}$ M$\odot$. However the first has a velocity (622 \kms) 
close to the HIJASS limit and NGC 4298 is confused with NGC 4302 (detected in 
HIJASS with $M_{HI}=10^{9.2}$ M$\odot$). Altogether the two mass estimates appear consistent with eachother, with the
exception of a few points at $M_{HI}<10^{8}$ M$\odot$ that appear slightly underestimated by D04. 
 
\subsection{The Virgo \HI\ mass function (HIMF)}

\begin{figure}
\centering
\includegraphics[width=9.0cm]{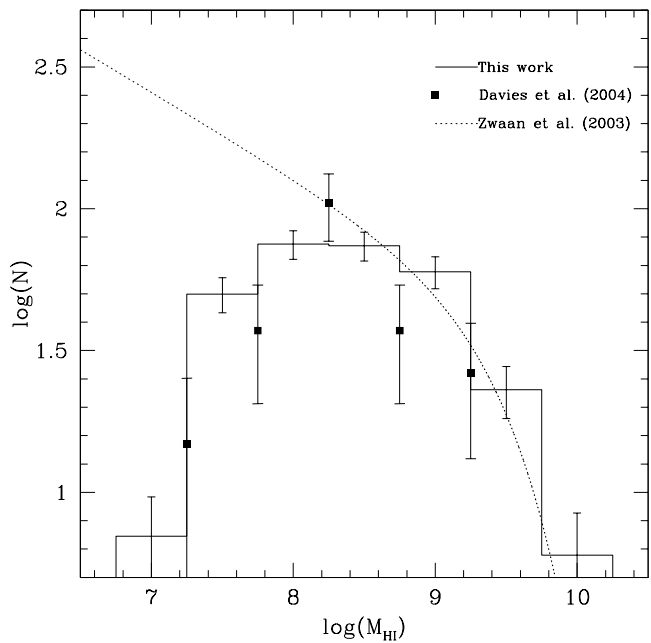}
\caption{The Virgo cluster \HI\ mass function (HIMF) as derived from this work using only the detected galaxies (histogram), together with
 the HIMF derived from the blind HIJASS survey (D04), normalized to our data 
as described in the text. Error bars represent purely statistical errors.
For comparison, the dotted line gives the Field
HIMF as derived by Zwaan et al. (2003), normalized arbitrarily in order to match our data at $M_{HI}=10^{9.5}$ M$\odot$.}
\label{HIfun1}
\end{figure}
\begin{figure}
\centering
\includegraphics[width=9.0cm]{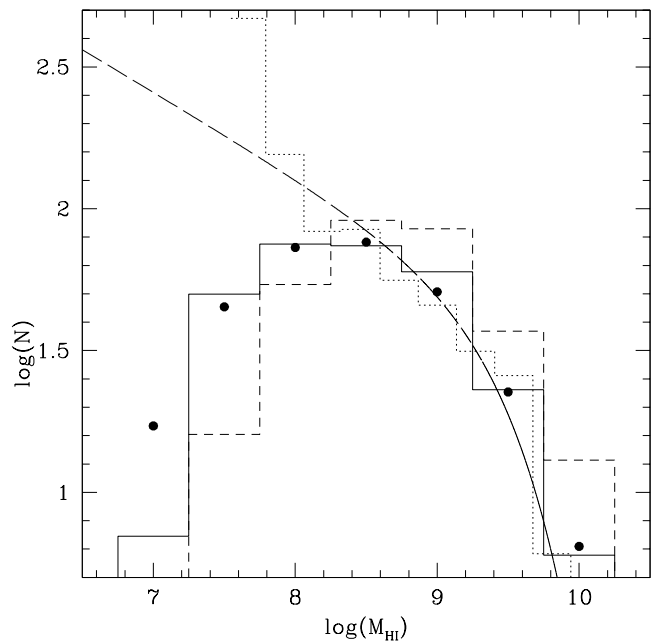}
\caption{The dotted histogram is obtained from the late-type galaxies 
optical luminosity function, as determined in the field by Marzke et al. (1998), transformed into $M_{HI}$
using the relation $log~M_{HI}=2.9-0.34\times M_p$ discussed in the text. It appears consistent with the field HILF
of Zwaan et al. (2003) (long dashed line) indicating that most of the HI is contributed to by late-type galaxies.
The dashed histogram is obtained from the late-type galaxies 
optical luminosity function of Virgo (see Fig \ref{optfun}), transformed into $M_{HI}$ using the same $M_{HI}$ vs. $M_p$ 
relation. The continuum histogram is the HIMF actually observed in Virgo (see Fig. \ref{HIfun1}). The filled dots
represent the Montecarlo simulation described in the text aimed at modeling the effects of the 
\HI\ deficiency on the HIMF.}
\label{HIfun2}
\end{figure}
\begin{figure}
\centering
\includegraphics[width=9.0cm]{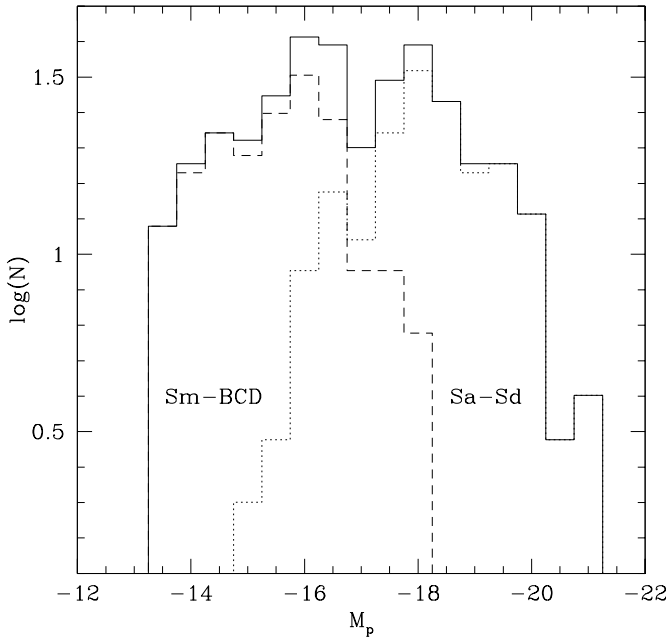}
\caption{The optical luminosity distribution of the 355 late-type galaxies analyzed in this work,
divided in Giants and Dwarfs, consistent with the one in Sandage et al. (1985) for $m_p$ $\leq$ $18.0$ mag,
accounting for the different assumed distance.}
\label{optfun}
\end{figure}

The data of our optically selected Virgo cluster galaxy sample as listed in Table \ref{sample_dat}, 
disregarding the undetected galaxies, were binned in $log(M_{HI})=0.5$ intervals, 
i.e. $\sim 2.5$ times the estimated uncertainty on log(\MHI) and were used to construct
the Virgo HIMF shown in Fig. \ref{HIfun1} (solid histogram) and to compare it with
the HIJASS HIMF of D04 (filled squares).
The latter was normalized to our data by the
ratio of the areas covered by the two surveys (a factor of 4.5) and by the ratio (a factor of 1.7) 
of the number of galaxies 
in the velocity interval $-500<V<3000$ \kms ~surveyed by us and those in the interval $500<V<2500$ \kms ~surveyed by
D04. 
Moreover the HIMF of D04 was shifted toward higher masses by $logM_{HI}=0.05$
to account for the slightly larger distances assumed by us (see Sect. \ref{daviescomp}). 
In addition, Fig. \ref{HIfun1} shows the Field
HIMF of Zwaan et al. (2003) arbitrarily normalized to our data.
In spite of the different construction methods  
the two Virgo HIMF are surprisingly consistent with eachother
\footnote{The point at $M_{HI}=10^{8.25}$ M$\odot$ of D04
includes 14 detections, 2 of which correspond to isolated \HI\ clouds -- disregarding
these two objects the discrepancy with the optical selected HIMF becomes negligible.}.
Both HIMFs show a maximum at $M_{HI}\sim 10^{8.5}$ M$\odot$ and a consistent negative slope for lower masses.
Below $M_{HI}\sim10^{8.5}$ M$\odot$ the two are inconsistent with the field HIMF which follows the slope +1.3.\\
The consistency between the Virgo radio- and optically- selected HIMFs in Fig.\ref{HIfun1} 
is not obvious. It implies that the contribution from isolated \HI\ clouds is negligible. 
Extrapolating from the 2 confirmed detections of D04 to the  7.5 times larger area covered
in the present survey, only 15 such objects are expected in the whole cluster
(that are missed by an optically selected \HI\ survey).
It also implies that,
besides the isolated \HI\ clouds, the bulk of the \HI\ emission is associated with the late-type
galaxies, with a negligible contribution from the early-type objects that we did not survey in 
\HI\ \footnote{A dozen \HI\ detections were reported in Virgo associated with early-type galaxies.
These have not however been surveyed with completeness. None belongs to the D04 sample.}.
We show in Fig. \ref{HIfun2} that also for the field the observed HIMF can be obtained purely from 
the contribution of the late-type population.
Using the relation $log~M_{HI}=2.9-0.34\times M_p$, which holds on average between the \HI\ mass
and the optical luminosity in a population of isolated, unperturbed galaxies (taken from GOLDmine), 
we transform the optical luminosity function of field S+Im galaxies by Marzke et al. (1998)
into an $\rm HILF_{S+Im}$, as shown by the dotted histogram of Fig. \ref{HIfun2}.
The actual measured HILF obtained by Zwaan et al. (2003) is consistent with the $\rm HILF_{S+Im}$, 
at least for $M_{HI}>10^{7.5}$ M$\odot$.\\
Similarly we transform the optical luminosity function of Virgo (Fig. \ref{optfun})
into the
$\rm HILF_{S+Im}$ (dashed histogram in Fig. \ref{HIfun2}) and compare it to the measured HILF
(continuum histogram).
The two are in general agreement, with some noticeable differences:
the $\rm HILF_{S+Im}$ is in excess over the measured HILF in the range $M_{HI}>10^{9}$ M$\odot$, 
while it lies below the measured HILF for $M_{HI}<10^{8}$ M$\odot$.
Both discrepancies can be understood in terms of \HI\ deficiency. 
Massive spirals in Virgo have large $Def_{HI}$ parameters that shift their measured $M_{HI}$ 
one or two decades below the corresponding values for isolated spirals, 
producing the measured HILF excess over the $\rm HILF_{S+Im}$ at $M_{HI}<10^{8}$ M$\odot$.
Solanes et al. (2001) found that the distribution of HI deficiency among 
galaxies of latest types in the Virgo cluster is skewed toward high values.
This can be either due to a real higher than average gas depletion or to the poorly
calibrated deficiency parameter for these systems, as mentioned earlier.
However Hoffman et al. (1985) confirm that the HI (hybrid) surface brightness is
monotonously decreasing toward later Hubble types and fainter optical luminosities. 
Disregarding these second order dependences of the HI deficiency parameter on the
Hubble type at extreme low luminosities, 
we have simulated the first order effect of the HI deficiency on the HIMF
by running a Montecarlo simulation with the
simple assumption that the $Def_{HI}$ parameter of galaxies in the Virgo cluster is Gaussian distributed
with a mean of 0.4 and a FWHM of 0.8, independent of the galaxy luminosity. 
Starting from the optical luminosity function
of S+Im of Fig.\ref{optfun} and assuming the $log~M_{HI}=2.9-0.34\times M_p$ relationship we have been
able to reproduce (filled dots) the observed HIMF (continuum histogram) of Fig.\ref{HIfun2}.

\subsection{The HIMF adopting upper limits}

\begin{figure}
\centering
\includegraphics[width=9.0cm]{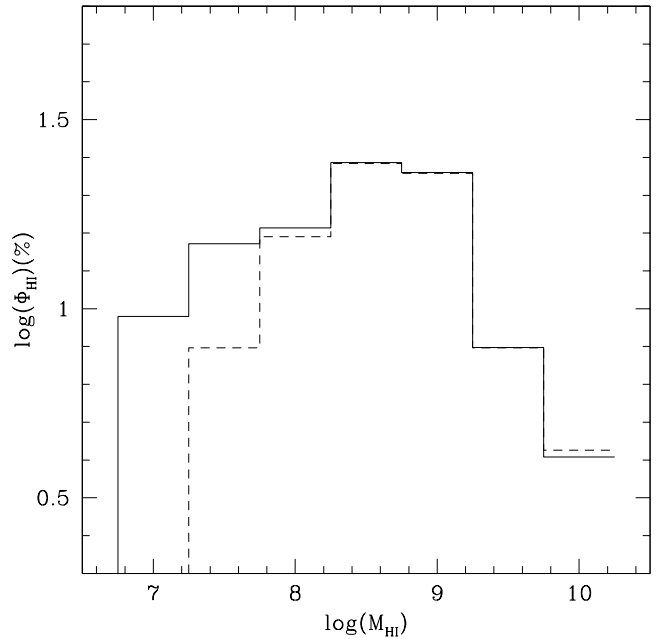}
\caption{
The fractional HI mass function as derived from the detected objects alone (dashed histogram) 
and including undetected galaxies (continuum histogram).}
\label{phi}
\end{figure}

So far we have shown that the HI mass function obtained from HI follow-up observations of 
an optically selected sample of late-type galaxies is in agreement with the HIMF derived 
from a blind HI survey. We have however not considered the relatively minor contribution from the
upper limits, i.e. 59/355 galaxies that were surveyed but not detected.\\
Radio astronomers have developed a robust method to account for upper limits when deriving the
continuum radio luminosity function (Avni et al. 1980; see a recent application to Virgo in Gavazzi
\& Boselli 1999) that we apply to the HI data.
Fig. \ref{phi} shows that the fractional HIMFs derived with (continuum histogram) and without (dashed
histogram) taking into account the contribution from upper limits are identical above $log~M_{HI}=8$M$\odot$.
Upper limits contribute for smaller masses, and make the faint-end slope of the HIMF less steep.
This is a small difference that should however not be disregarded. 

\section{Conclusions}

We have observed in the 21-cm \HI\ line, with the refurbished Arecibo telescope, 33 galaxies in the Virgo cluster. 
Given the high sensitivity of our observations (rms noise $\sim 0.5$ mJy
corresponding to $log~M_{HI}=6.6$$\rm M\odot$ )
12 objects were detected and stringent upper limits were obtained for the remaining ones. \\
In the Virgo area covered by the VCC the new observations brought to 100 \% completeness the
\HI\ survey of late-type galaxies with $m_p \leq 18.0$ mag which are cluster members or bona-fide members.\\
Using the Virgo \HI\ data set, comprising 355 late-type galaxies (296 of which are positive detections) 
we construct the cluster \HI\ mass function (HIMF) as derived from an optically selected 
\HI\ survey. Considering or disregarding the contribution from HI non-detections,
we find it in remarkable agreement with the radio selected HIMF available for this cluster
from Davies et al. (2004).\\
The low-mass end of the Virgo HIMF is inconsistent with the field HIMF of Zwaan et al. (2003).\\
We show that both the Virgo and the field HIMFs can be obtained from the optical luminosity
function of S+Im galaxies alone, under the assumption that $M_{HI}$ scales with $M_p$ 
according to the universal relation: $log~M_{HI}=2.9-0.34\times M_p$.\\
The latter evidence allows us to conclude that  
neutral hydrogen in the local universe is primarily contributed by late-type galaxies, 
with marginal contributions from early-type galaxies and isolated \HI\ clouds.\\
The inconsistency between the cluster and the field HIMF
derives primarily from the difference in the optical luminosity function of late-type galaxies 
in the two environments and from the HI deficiency occurring in spirals in rich clusters.

\acknowledgements{ 
We thank Luca Cortese for running the Montecarlo simulations and for useful discussions. 
The Arecibo Observatory is part of the National Astronomy and Ionosphere Center, 
which is operated by Cornell University under a cooperative agreement with the 
National Science Foundation. This research also has made use of the Goldmine database, of the Lyon-Meudon 
Extragalactic Database (Leda), recently incorporated in HyperLeda and of the NASA/IPAC 
Extragalactic Database (NED) which is operated by the Jet Propulsion Laboratory, 
California Institute of Technology, under contract with the National Aeronautics and Space      
Administration.

\newpage
   \onecolumn

\setcounter{table}{1}
\begin{table*}
{\footnotesize
\begin{tabular}{llrrrrrrrlc}
\multicolumn{11}{l}{\footnotesize {\bf Table 1} Parameters of the newly observed galaxies.} \\
\smallskip \\
\hline
\vspace{-2 mm} \\
Obj. & RA & Dec & $V_{opt}$ & $\sigma$ & $S_p$ & $V_{HI}$ & $W_{50}$ & $W_{20}$ & $I_{HI}$ & Qual. \\ 
 & \multicolumn{2}{c}{(J2000.0)} & (km/s) & (mJy) & (mJy) & (km/s) & (km/s) & (km/s) & (Jy km/s) \\   
\hline
\vspace{-2 mm} \\
\multicolumn{10}{l}{Virgo Cluster} \\ 
\vspace{-2 mm} \\
VCC 1        & 120820.02 & 134100.2 & 2275$\pm$43  & 0.34 &  1.6 &  2240$\pm$5  & 107 & 124 & 0.13$\pm$0.03  & 2\\ 
VCC 48       & 121215.11 & 122917.8 &  -52$\pm$60  & 0.64 &   -  &   -          &  -  &  -  &   -  & - \\ 
VCC 99       & 121402.18 & 064323.3 & 2454$\pm$44  & 0.72 &  6.1 &  2418$\pm$4  & 169 & 189 & 0.72$\pm$0.08  & 1  \\ 
VCC 222      & 121709.83 & 071129.1 & 2298$\pm$122 & 0.60 &   -  &   -          &  -  &  -  &   - & - \\ 
VCC 227      & 121714.38 & 085632.1 & 1290$\pm$60  & 0.34 &  1.2 & 1304:$\pm$7  &  24 &  41 & 0.02$\pm$0.02& 4 \\ 
VCC 256      & 121747.74 & 042838.3 & 2103$\pm$60  & 0.44 &  1.7 & 2007:$\pm$40 & 307 & 410 & 0.31$\pm$0.07  & 4 \\ 
VCC 275      & 121811.29 & 093002.4 & 1733$\pm$60  & 0.54 &   -  &   -          &  -  &  -  &   - & -  \\ 
VCC 315      & 121900.46 & 060540.7 & 1594$\pm$46  & 0.57 &   -  &   -          &  -  &  -  &   - & -  \\ 
VCC 323      & 121906.39 & 054332.7 & 2759$\pm$35  & 0.53 &   -  &   -          &  -  &  -  &   - & -  \\ 
VCC 341      & 121922.15 & 060554.8 & 1827$\pm$60  & 0.76 &   -  &   -          &  -  &  -  &   - & -  \\ 
VCC 358      & 121935.66 & 055047.9 & 2599$\pm$59  &   28 &   -  &   -          &  -  &  -  &   - & -  \\ 
VCC 362      & 121942.19 & 053216.9 &  156$\pm$44  & 0.44 &   -  &   -          &  -  &  -  &   - & -  \\
VCC 517      & 122201.37 & 050603.8 & 1888$\pm$51  & 0.37 &  3.4 &  1864$\pm$12 & 285 & 356 & 0.37$\pm$0.06  & 1\\ 
VCC 524      & 122205.74 & 090238.8 & 1117$\pm$98  & 0.51 &  5.1 &  1035$\pm$1  & 322 & 329 & 1.14$\pm$0.07  & 1\\ 
VCC 528      & 122207.76 & 060611.8 & 7019$\pm$60  & 0.63 &  5.2 &  7046$\pm$9  & 261 & 309 & 0.86$\pm$0.09  & 1\\ 
VCC 531      & 122210.88 & 045705.9 & 1912$\pm$60  & 0.56 &   -  &   -          &  -  &  -  &   -  & -\\ 
VCC 666      & 122346.13 & 164728.5 &         -    & 0.51 &   -  &   -          &  -  &  -  &   -  &  -\\ 
VCC 675      & 122354.35 & 030504.6 & 1860$\pm$42  & 0.78 & 14.0 &  1857$\pm$   &  91 & 128 & 1.14$\pm$0.07  & 2\\ 
VCC 679      & 122355.17 & 112928.6 &         -    & 1.19 &   -  &   -          &  -  &  -  &   -  & - \\ 
VCC 802      & 122529.01 & 132947.3 & -215$\pm$42  & 0.58 &   -  &   -          &  -  &  -  &   -  & - \\ 
VCC 1086     & 122816.00 & 092610.6 &  294$\pm$39  & 0.63 &  6.7 &  328$\pm$8   & 240 & 294 & 0.93$\pm$0.09  & 3 \\ 
VCC 1121     & 122841.73 & 110754.9 &         -    & 0.66 &   -  &   -          &  -  &  -  &   -  & -  \\ 
VCC 1189     & 122928.83 & 064612.3 &  544$\pm$43  & 1.17 & 49.5 &  516$\pm$1   & 112 & 132 & 4.17$\pm$0.11  & 1\\ 
VCC 1196     & 122931.25 & 140258.3 &  908$\pm$36  & 0.57 &   -  &   -          &  -  &  -  &   -  & - \\ 
VCC 1237     & 122951.15 & 135203.5 & -335$\pm$60  & 0.77 &   -  &   -          &  -  &  -  &   -  & - \\ 
VCC 1287     & 123023.79 & 135855.8 &         -    & 0.72 &   -  &   -          &  -  &  -  &   -  & - \\ 
VCC 1358     & 123122.99 & 171223.3 &         -    & 0.66 &   -  &   -          &  -  &  -  &   -  & - \\ 
VCC 1377     & 123139.21 & 105008.5 &         -    & 0.51 &   -  &   -          &  -  &  -  &   -  & - \\ 
VCC 1435     & 123232.42 & 080239.0 &  609$\pm$42  & 0.44 &   -  &   -          &  -  &  -  &   -  & - \\ 
VCC 1448     & 123240.83 & 124613.1 &         -    & 0.92 &   -  &   -          &  -  &  -  &   -  & - \\ 
VCC 1597     & 123502.77 & 052534.6 &  861$\pm$60  & 0.39 &  5.4 &   841$\pm$2  & 124 & 144 & 0.51$\pm$0.04  & 1\\ 
VCC 1885     & 124137.57 & 154933.2 &         -    & 0.38 &   -  &   -          &  -  &  -  &   -  & - \\ 
VCC 1970     & 124329.11 & 100534.7 & 1325$\pm$60  & 0.37 &  3.0 &  1324$\pm$12 &  70 & 136 & 0.13$\pm$0.03 & 3\\ 
\vspace{-2 mm} \\																			    
\hline
\end{tabular}
}
\end{table*}
\newpage
   \scriptsize{
   \begin{longtable}{rcccclrcrc}
   \caption{Basic \HI\ properties of late-type Virgo galaxies.}\\
   \hline
   \hline
   \noalign{\smallskip}
    VCC & Type & $m_p$ & a & Cloud & V & log $M_{HI}$ &  $Def_{HI}$ & Qual & Ref.  \\  
             &      & (mag) & (arcmin) & &  \kms & ($M\odot$)   &      &            &  \\
       (1)   &  (2) & (3) & (4) & (5) & (6)   &  (7)  &  (8)    & (9) & (10) \\	     
   \noalign{\smallskip}
   \hline
   \noalign{\smallskip}
   \endfirsthead
   \caption{Continue}\\
   \hline
   \noalign{\smallskip}
    VCC & Type & $m_p$ & a & Cloud & V & log $M_{HI}$ &  $Def_{HI}$ & Qual & Ref.  \\  
             &      & (mag) & (arcmin) & &  \kms & ($M\odot$)   &      &            &  \\
	        \noalign{\smallskip}
   \hline
   \noalign{\smallskip}
   \endhead
   \hline
   \endfoot          
     1  &      BCD  & 14.78  &   0.80 & M &	2240 &        7.50  &	1.14  &    2 &  T.W.	  \\
     4  &      Im   & 17.50  &   0.50 & M &	 589 &        8.25  &	0.00  &    2 &  HH87	  \\
    10  &      BCD  & 14.75  &   1.03 & M &	1973 &        8.77  &	0.08  &    1 &  HH87	  \\
    15  &      Sa   & 14.70  &   1.37 & M &	2545 &        8.55  &	0.43  &    1 &  HL89	  \\
    17  &      Im   & 15.20  &   0.91 & M &	 819 &        8.78  &  -0.04  &    2 &  HH87	  \\
    22  &      BCD  & 16.00  &   0.27 & M &	1695 &        8.23  &  -0.49  &    2 &  HB03	  \\
    24  &      BCD  & 14.95  &   1.00 & M &	1292 &        8.96  &  -0.15  &    1 &  HB03	  \\
    25  &      Sc   & 12.46  &   2.54 & M &	2169 &        9.73  &  -0.21  &    1 &   M94	  \\
    26  &      Im   & 17.50  &   0.43 & M &	2469 &        8.38  &  -0.26  &    3 &  HH87	  \\
    34  &      Sc   & 14.65  &   1.16 & M &	 266 &        8.99  &  -0.12  &    1 &  HL89	  \\
    47  &      Sa   & 14.20  &   1.41 & M &	1862 &        8.39  &	0.61  &    1 &  HG86	  \\
    48  &      Sdm  & 14.30  &   1.71 & M &	 -52 &$<$     7.29  &	1.97  &    - &  T.W.	  \\
    52  &      Im   & 17.80  &   0.39 & W &	2088 &        7.64  &	0.42  &    5 &  HH87	  \\
    58  &      Sb   & 13.17  &   2.54 & M &	2207 &        9.46  &	0.11  &    1 &  HH84	  \\
    66  &      Sc   & 11.89  &   5.35 & N &	 369 &        9.81  &  -0.20  &    1 &  W86	  \\
    67  &      Sc   & 13.98  &   2.26 & M &	-183 &        9.17  &	0.25  &    1 &  HL89	  \\
    73  &      Sb   & 13.35  &   1.89 & W &	2082 &        8.96  &	0.40  &    2 &  HL89	  \\
    74  &      BCD  & 16.30  &   0.85 & N &	 861 &$<$     7.02  &	1.15  &    - &  HH87,D04* \\
    79  &      Im   & 17.20  &   0.57 & N &	   - &$<$     7.12  &	0.73  &    - &  HH87,D04  \\
    81  &      Sc   & 15.60  &   0.95 & N &	2075 &        8.63  &  -0.45  &    1 &  HG86,D04* \\ 
    83  &      Im   & 15.13  &   1.26 & N &	2439 &        8.04  &	0.46  &    1 &  HH87,D04* \\ 
    87  &      Sm   & 15.00  &   1.45 & N &	-134 &        8.32  &	0.29  &    2 &  HH87,D04  \\
    89  &      Sc   & 12.53  &   2.26 & M &	2116 &        9.46  &  -0.03  &    1 &  HH84,D04* \\ 
    92  &      Sb   & 10.92  &   9.78 & N &	-135 &        9.76  &	0.33  &    1 &  HR89,D04  \\ 
    97  &      Sc   & 13.20  &   1.96 & M &	2476 &        9.08  &	0.23  &    1 &   M94,D04* \\ 
    99  &      Sa   & 14.81  &   1.41 & W &	2418 &        8.24  &	0.77  &    1 &  T.W.	  \\
   105  &      Sd   & 13.68  &   2.48 & W &	1221 &        9.27  &	0.29  &    1 &  HL89	  \\
   114  &      Im   & 16.00  &   0.64 & W &	2071 &        8.70  &  -0.24  &    2 &  HH87	  \\
   117  &      Im   & 16.50  &   0.64 & W &	1788 &        8.93  &  -0.47  &    1 &  HH87	  \\
   119  &      Sc   & 14.76  &   1.71 & M &	 620 &        9.04  &	0.15  &    1 &  HL89,D04* \\
   120  &      Scd  & 13.47  &   3.60 & W &	2064 &        9.70  &	0.17  &    1 &  HH84	  \\
   124  &      Sm   & 16.00  &   0.71 & M &	2084 &        7.73  &	0.81  &    3 &  HH87,D04* \\
   126  &      Sd   & 14.42  &   1.87 & N &	 263 &        8.50  &	0.32  &    1 &  HL89,D04  \\
   130  &      BCD  & 16.50  &   0.63 & N &	2189 &        7.86  &	0.06  &    1 &  HH87	  \\
   131  &      Sc   & 14.34  &   2.60 & N &	2317 &        8.93  &	0.09  &    1 &  HG86,D04* \\ 
   132  &      Sd   & 16.40  &   1.36 & N &	2085 &        8.11  &	0.44  &    2 &  HL89,D04* \\ 
   135  &    S/BCD  & 14.81  &   1.16 & M &	2378 &$<$     7.19  &	1.75  &    - &  HG86,D04* \\
   143  &      Sc   & 15.46  &   1.00 & N &	 375 &        7.88  &	0.35  &    1 &  HL89,D04  \\
   144  &      BCD  & 15.31  &   0.63 & W &	2014 &        8.74  &  -0.31  &    2 &  HH87	  \\
   145  &      Sc   & 12.77  &   5.10 & N &	 702 &        9.39  &	0.19  &    1 &  HH84,D04* \\ 
   152  &      Scd  & 13.48  &   1.96 & N &	 592 &        8.61  &	0.24  &    1 &  HL89	  \\ 
   157  &      Sc   & 11.50  &   3.60 & N &	 -83 &        8.68  &	0.61  &    1 &  HH84,D04  \\
   159  &      Im   & 15.08  &   1.04 & W &	2584 &        8.55  &	0.30  &    2 &  HH87	  \\
   162  &      Sd   & 14.41  &   2.92 & N &	1979 &        8.88  &	0.30  &    1 &  HL89	  \\
   167  &      Sb   & 10.97  &   9.12 & N &	 140 &        9.35  &	0.69  &    1 &  CG90,D04  \\
   168  &      Im   & 17.10  &   0.43 & N &	 682 &        7.35  &	0.26  &    2 &  HH87,D04* \\
   169  &      Im   & 16.50  &   0.85 & N &	2222 &        8.52  &  -0.35  &    2 &  HH87	  \\
   170  &      Sd   & 14.56  &   1.16 & N &	1411 &        7.45  &	0.98  &    2 &  HG86,D04* \\
   171  &      Im   & 17.40  &   0.57 & W &	 875 &        7.16  &	1.20  &    4 &  HW89	  \\
   172  &      BCD  & 14.50  &   1.26 & W &	2175 &        9.06  &  -0.05  &    - &  HH87	  \\
   187  &      Scd  & 13.91  &   3.52 & N &	 226 &        8.93  &	0.40  &    1 &  HH84,D04  \\
   199  &      Sa   & 12.95  &   2.92 & W &	2594 &        8.81  &	0.72  &    1 &   M94	  \\
   207  &      BCD  & 17.20  &   0.36 & W &	2564 &        8.25  &  -0.25  &    2 &  HW89	  \\
   213  &    S/BCD  & 14.26  &   0.93 & N &	-162 &        8.00  &	0.25  &    2 &  HG86,D04  \\
   217  &      Im   & 15.50  &   1.71 & N &	1183 &        8.61  &	0.14  &    2 &  HG86	  \\
   221  &      Sc   & 13.43  &   1.76 & W &	2031 &        8.81  &	0.41  &    2 &  HL89	  \\
   222  &      Sa   & 12.62  &   4.33 & W &	2298 &$<$     7.81  &	1.99  &    - &  T.W.	  \\
   223  &      BCD  & 16.50  &   0.34 & W &	2070 &        8.22  &  -0.29  &    2 &  HH87	  \\
   224	&      Scd  & 14.70  &   1.87 & N &	2133 &        8.62  &	0.20  &    1 &  HG86,D04* \\ 
   226  &      Sc   & 12.53  &   2.01 & N &	 864 &        8.32  &	0.48  &    1 &   M94,D04* \\ 
   227  &      Sdm  & 14.90  &   1.16 & W &	1304 &        6.68  &	2.25  &    3 &  T.W.	  \\
   234  &      Sa   & 12.99  &   3.36 & W &	2237 &        8.37  &	1.26  &    1 &   M94	  \\
   241  &      Sd   & 14.60  &   2.60 & N &	-163 &        8.72  &	0.36  &    1 &  HG86,D04  \\
   260  &      Im   & 15.70  &   1.03 & W &	1775 &        8.03  &	0.82  &    2 &  HH87	  \\
   267  &      Sbc  & 13.82  &   2.01 & B &	 733 &        9.04  &	0.15  &    1 &  HL89	  \\
   274  &      BCD  & 17.50  &   0.57 & W &	   - &$<$     7.49  &	0.87  &    - &  HW89	  \\
   275  &      Im   & 14.54  &   1.79 & W &	1733 &$<$     7.22  &	2.08  &    - &  T.W.	  \\
   280  &      Im   & 17.70  &   0.28 & N &	   - &$<$     6.89  &	0.38  &    - &  HH87	  \\
   281  &    S/BCD  & 15.38  &   0.14 & N &	 257 &        7.58  &  -0.91  &    2 &  HH87,D04  \\
   286  &      Im   & 16.00  &   0.51 & W &	1822 &        7.93  &	0.35  &    3 &  HH87	  \\
   289  &      Sc   & 14.81  &   1.71 & W &	 863 &        8.89  &	0.30  &    1 &  HL89	  \\
   297	&      Sc   & 15.10  &   1.16 & B &	1999 &        8.27  &	0.32  &    1 &  HL95	  \\
   307  &      Sc   & 10.43  &   6.15 & N &	2405 &        9.71  &	0.01  &    1 &  HS81,D04* \\ 
   309  &    Im/BCD & 16.20  &   0.64 & N &	1566 &        7.71  &	0.23  &    2 &  HH87,D04* \\
   313	&      Sa   & 14.62  &   1.26 & W &	2376 &        8.78  &	0.14  &      &  HR86	     3 \\
   315  &      Sa   & 14.98  &   1.10 & W &	1594 &$<$     7.79  &	1.04  &    - &  T.W.	  \\
   318  &      Scd  & 14.01  &   1.71 & W &	2469 &        9.39  &  -0.13  &    1 &  HL89	  \\
   322  &      Im   & 15.10  &   1.26 & N &	-206 &        8.25  &	0.25  &    2 &  HG86,D04  \\
   323  &      Sa   & 14.91  &   1.16 & W &	2759 &$<$     7.76  &	1.10  &    - &  T.W.	  \\
   324  &      BCD  & 14.78  &   1.35 & S &	1524 &        8.19  &	0.36  &    2 &  HH87	  \\
   328  &      Im   & 16.90  &   1.00 & N &	2179 &        7.98  &	0.33  &    2 &  HG86,D04* \\
   329  &      Im   & 16.80  &   0.63 & B &	1622 &        7.98  &	0.19  &    2 &  HH87	  \\
   331  &     Pec   & 15.00  &   1.00 & W &	1984 &        8.13  &	0.89  &    3 &  HL95	  \\
   334  &      BCD  & 15.87  &   0.56 & N &	-254 &        7.95  &  -0.11  &    2 &  HH87,D04  \\
   340  &      BCD  & 14.43  &   1.10 & W &	1512 &        8.82  &	0.08  &    2 &  HH87	  \\
   341  &      Sa   & 12.70  &   3.52 & B &	1827 &$<$     7.63  &	1.79  &    - &  T.W.	  \\
   343  &      Sd   & 15.10  &   1.10 & B &	2479 &        7.98  &	0.65  &    2 &  HL89	  \\
   350  &      Im   & 17.05  &   0.66 & N &	 305 &        7.97  &  -0.01  &    3 &  HH87,D04  \\
   358  &      Sa   & 13.80  &   1.55 & B &	2599 &$<$     8.65  &	0.18  &    - &  HR86	  \\
   362  &      Sa   & 14.51  &   2.16 & W &	 156 &$<$     7.68  &	1.63  &    - &  T.W.	  \\
   364  &      Im   & 17.30  &   0.74 & N &	   - &$<$     7.05  &	1.00  &    - &  HH87,D04  \\
   367  &      Im   & 17.20  &   0.56 & W &	2350 &        7.99  &	0.37  &    1 &  T.W.	  \\
   381  &      Im   & 16.50  &   0.85 & B &	 482 &        8.25  &	0.17  &    2 &  HH87	  \\
   382  &      Sc   & 12.37  &   2.01 & W &	2378 &        9.65  &  -0.32  &    1 &  HH84	  \\
   386  &      Sa   & 14.47  &   1.55 & W &	2380 &        9.25  &  -0.18  &    1 &  DR00	  \\
   393  &      Sc   & 13.25  &   2.10 & B &	2617 &        8.76  &	0.33  &    1 &  HL89	  \\
   404  &      Scd  & 15.00  &   1.71 & S &	1733 &        8.34  &	0.41  &    1 &  HL89	  \\
   410  &      BCD  & 17.10  &   0.31 & N &	 284 &        7.38  &  -0.03  &    3 &  HH87,D04  \\
   415  &      Sd   & 14.82  &   1.16 & B &	2560 &        8.28  &	0.39  &    1 &  HL89	  \\
   423  &      Im   & 17.30  &   0.51 & S &	2384 &        8.06  &  -0.30  &    1 &  HW89	  \\
   425  &      Im   & 17.30  &   0.43 & B &	   - &$<$     6.98  &	0.87  &    - &  HW89	  \\
   428  &      BCD  & 17.50  &   0.39 & A &	 794 &        7.63  &  -0.08  &    2 &  HH87,D04* \\
   446  &    Im/BCD & 15.50  &   0.85 & B &	 825 &        7.68  &	0.74  &    3 &  HH87	  \\
   448  &      Im   & 16.80  &   0.39 & A &	 672 &        7.79  &  -0.25  &    2 &  HH87,D04* \\
   449  &      Sbc  & 14.34  &   4.33 & S &	2541 &        9.02  &	0.51  &    1 &  HL89	  \\
   453  &      Sm   & 16.00  &   0.79 & N &	 910 &        7.71  &	0.41  &    2 &  HH87	  \\
   459  &      BCD  & 14.95  &   0.84 & A &	2108 &        8.22  &  -0.07  &    2 &  HH87,D04* \\ 
   460  &      Sa   & 11.20  &   5.10 & A &	 921 &        7.62  &	1.85  &    3 &   M94,D04* \\
   464  &      BCD  & 17.50  &   0.64 & B &	   - &$<$     7.05  &	1.14  &    - &  HW89	  \\
   465  &      Sc   & 12.62  &   3.95 & N &	 357 &        9.26  &	0.10  &    1 &  HH84	  \\
   467  &      Im   & 17.70  &   0.43 & S &	2435 &        7.58  &	0.03  &    1 &  HW89	  \\
   468  &      BCD  & 16.00  &   0.56 & S &	1980 &        7.68  &	0.16  &    2 &  HB03	  \\
   476  &      Im   & 17.90  &   0.36 & N &	   - &$<$     7.09  &	0.40  &    - &  HW89	  \\
   477  &      Im   & 16.96  &   1.00 & A &	1866 &        7.53  &	0.77  &    3 &  HH87,D04* \\
   483  &      Sc   & 12.08  &   3.60 & A &	1136 &        8.94  &	0.34  &    3 &  HH84,D04* \\
   491  &      Scd  & 12.86  &   1.96 & N &	 234 &        9.14  &  -0.29  &    1 &  HH84	  \\
   492  &      Sa   & 13.76  &   2.16 & B &	2310 &$<$     7.79  &	1.28  &    - &  HL89	  \\
   497  &      Sc   & 12.55  &   6.74 & A &	1150 &        9.24  &	0.56  &    1 &  HH84,D04* \\ 
   508  &      Sc   & 10.17  &   6.59 & S &	1568 &        9.85  &  -0.06  &    1 &  HH84	  \\
   509  &      Sdm  & 14.98  &   1.45 & B &	1258 &        8.47  &	0.38  &    1 &  HH87	  \\
   512  &      Sm   & 15.69  &   1.45 & A &	 153 &        8.40  &	0.21  &    1 &  HH87	  \\
   513  &      BCD  & 15.10  &   0.73 & S &	1832 &        7.27  &	0.79  &    2 &  HH87	  \\
   514  &      Sc   & 14.70  &   1.41 & B &	 851 &        8.12  &	0.65  &    2 &  HL89	  \\
   517  &      Sab  & 14.90  &   1.00 & S &	1864 &        7.40  &	0.91  &    1 &  T.W.	  \\
   520  &      Im   & 17.50  &   0.50 & B &	   - &$<$     7.43  &	0.55  &    - &  HW89	  \\
   522  &      Sa   & 13.19  &   2.60 & A &	1888 &$<$     7.44  &	1.55  &    - &  GK83,D04* \\
   524  &      Sbc  & 12.79  &   3.95 & B &	1035 &        8.15  &	1.53  &    1 &  T.W.	  \\
   530  &      Im   & 15.80  &   1.29 & A &	1297 &        7.60  &	0.91  &    2 &  HH87,D04* \\
   531  &      Sa   & 15.00  &   1.10 & S &	1912 &$<$     7.24  &	1.14  &    - &  T.W.	  \\
   534  &      Sa   & 13.59  &   2.01 & B &	1071 &        7.64  &	1.38  &    3 &   M94	  \\
   552  &      Sc   & 13.61  &   1.89 & S &	1296 &        9.17  &  -0.42  &    1 &  HL89	  \\
   559  &      Sab  & 12.56  &   5.10 & A &	 153 &        8.09  &	1.38  &    1 &  HH84,D04  \\
   562  &      BCD  & 16.20  &   0.63 & A &	  44 &$<$     7.02  &	0.90  &    - &  HH87,D04  \\
   565  &      Im   & 15.70  &   0.93 & B &	 877 &        7.77  &	0.73  &    2 &  HH87	  \\
   566  &      Sm   & 15.80  &   0.71 & B &	1407 &        8.59  &  -0.32  &    2 &  HH87	  \\
   567  &      Scd  & 14.36  &   2.16 & B &	2366 &        8.82  &	0.36  &    2 &  HL89	  \\
   570  &      Sab  & 12.73  &   5.10 & A &	1443 &        8.11  &	1.36  &    1 &  HH84	  \\
   576  &      Sbc  & 13.70  &   2.48 & B &	1254 &        9.20  &	0.15  &    1 &  HG86	  \\
   583  &      Im   & 15.76  &   1.16 & A &	 -72 &$<$     7.12  &	1.31  &    - &  HH87,D04  \\
   584  &      Im   & 15.80  &   0.71 & B &	  56 &        7.00  &	1.27  &    5 &  HH87	  \\
   585  &      Im   & 17.00  &   1.16 & A &	   - &$<$     7.09  &	1.34  &    - &  HH87	  \\
   596  &      Sc   & 10.11  &   9.12 & A &	1575 &        9.52  &	0.53  &    1 &  HS81,D04* \\ 
   613  &      Sa   & 12.60  &   3.52 & S &	1670 &        8.83  &	0.38  &    1 &  HW89	  \\
   618  &      Im   & 16.50  &   0.60 & A &	1890 &        7.94  &  -0.05  &    2 &  HH87,D04* \\ 
   620  &      Sm   & 15.20  &   1.26 & A &	 746 &        8.02  &	0.48  &    2 &  HG86	  \\
   630  &      Sd   & 13.10  &   5.86 & A &	1564 &        8.59  &	1.16  &    1 &  HH84	  \\
   641  &      BCD  & 15.08  &   0.73 & B &	 906 &        8.15  &	0.15  &    3 &  HH87	  \\
   655  &    S/BCD  & 13.21  &   1.55 & A &	1147 &        7.91  &	0.75  &    2 &  GK83,D04* \\ 
   656  &      Sb   & 13.14  &   2.48 & B &	1014 &        8.79  &	0.53  &    1 &  HH84	  \\
   664  &      Sc   & 13.50  &   2.60 & A &	-427 &        8.40  &	0.62  &    2 &  HS81,D04  \\
   666  &      Im   & 16.80  &   1.00 & A &	   - &$<$     6.65  &	1.66  &    - &  T.W.,D04  \\
   667  &      Sc   & 14.24  &   1.71 & B &	1420 &        8.34  &	0.58  &    1 &  HL89	  \\
   675  &      Sa   & 15.00  &   0.56 & S &	1857 &        7.89  &	0.01  &    2 &  T.W.	  \\
   688  &      Sc   & 13.94  &   1.41 & B &	1125 &        8.32  &	0.45  &    1 &  HL89	  \\
   692  &      Sc   & 12.93  &   2.92 & A &	2324 &        8.46  &	0.66  &    1 &  HS81,D04* \\ 
   693  &      Sm   & 15.06  &   1.16 & S &	2048 &        8.10  &	0.32  &    1 &  HH87	  \\
   697  &      Sc   & 14.17  &   1.55 & B &	1231 &        8.22  &	0.62  &    2 &  HL89	  \\
   699  &     Pec   & 14.22  &   1.95 & B &	 727 &        9.03  &	0.19  &    2 &  HL89	  \\
   713  &      Sc   & 14.04  &   3.20 & B &	1137 &        8.10  &	1.34  &    3 &  HG86	  \\
   737  &    S/BCD  & 14.94  &   1.07 & S &	1725 &        8.46  &  -0.09  &    1 &  HH87	  \\
   739  &      Sd   & 14.37  &   2.01 & S &	 927 &        8.78  &	0.10  &    2 &  HL89	  \\
   740  &      Sm   & 15.70  &   0.71 & B &	 875 &        8.19  &	0.08  &    2 &  HH87	  \\
   741  &      BCD  & 15.50  &   0.84 & S &	1861 &        8.04  &	0.12  &    1 &  HH87	  \\
   768  &      Sc   & 14.91  &   1.03 & A &	2434 &        8.09  &	0.16  &    1 &  HL89,D04* \\
   772  &      BCD  & 17.00  &   0.51 & S &	1226 &        7.67  &	0.09  &    2 &  HH87	  \\
   785  &      Sa   & 12.16  &   3.06 & S &	2557 &        9.00  &	0.10  &    1 &  HH84	  \\
   787  &      Scd  & 13.69  &   1.84 & B &	1136 &        8.79  &	0.26  &    1 &  SS90	  \\
   792  &      Sab  & 12.36  &   3.52 & B &	 971 &        8.57  &	0.85  &    1 &  HH84	  \\
   793  &      Im   & 16.74  &   0.47 & A &	1906 &        7.65  &	0.05  &    2 &  HH87,D04* \\
   802  &      BCD  & 17.40  &   0.64 & A &	-215 &$<$     6.70  &	1.24  &    - &  T.W.,D04  \\
   809  &      Sc   & 14.55  &   1.45 & A &	-142 &        8.39  &	0.15  &    2 &  HG86,D04  \\
   825  &      Im   & 15.90  &   1.00 & B &	   - &$<$     7.16  &	1.40  &    - &  HH87	  \\
   826  &      Im   & 15.00  &   1.20 & S &	1507 &        8.32  &	0.14  &    2 &  HH87	  \\
   827  &      Sc   & 13.76  &   3.60 & B &	 992 &        9.45  &	0.08  &    1 &  HL89	  \\
   836  &      Sab  & 11.83  &   5.10 & A &	2515 &        8.78  &	0.69  &    1 &  HS81,D04* \\
   841  &      BCD  & 15.60  &   0.84 & A &	 501 &        7.61  &	0.55  &    2 &  HG86,D04* \\
   848  &    Im/BCD & 14.72  &   1.16 & B &	1537 &        8.85  &  -0.18  &    2 &  HH87	  \\
   849  &      Sbc  & 13.27  &   2.18 & B &	1103 &        8.84  &	0.41  &    1 &  SA82	  \\
   851  &      Sc   & 14.14  &   2.16 & B &	1195 &        8.88  &	0.23  &    1 &  HG86	  \\
   857  &      Sb   & 11.76  &   3.60 & A &	 914 &        8.51  &	0.86  &    1 &  HS81,D04* \\ 
   859  &      Sc   & 14.61  &   2.92 & S &	1428 &        8.62  &	0.49  &    1 &  HL89	  \\
   865  &      Sc   & 13.02  &   3.36 & A &	-124 &        8.85  &	0.38  &    1 &  SS90,D04  \\
   873  &      Sc   & 12.56  &   3.95 & A &	 234 &        8.74  &	0.63  &    1 &  HH84,D04  \\
   874  &      Sc   & 12.99  &   1.89 & A &	1738 &        7.81  &	0.95  &    3 &   M94,D04* \\ 
   888  &      Im   & 15.78  &   1.16 & B &	1090 &        8.25  &	0.41  &    2 &  HH87	  \\
   890  &      BCD  & 16.00  &   0.21 & B &	1483 &        7.33  &  -0.05  &    2 &  HW89	  \\
   905  &      Sc   & 13.42  &   2.79 & B &	1290 &        8.97  &	0.35  &    2 &  HL89	  \\
   912  &      Sbc  & 12.97  &   2.92 & A &	 105 &        8.26  &	0.99  &    1 &  HS81,D04  \\
   921  &      Sbc  & 13.14  &   1.89 & S &	2289 &        8.33  &	0.59  &    2 &  HL89	  \\
   938  &      Sc   & 13.28  &   2.18 & S &	1395 &        8.52  &	0.36  &    1 &  HH84	  \\
   939  &      Sc   & 12.92  &   3.45 & B &	1271 &        9.26  &	0.24  &    2 &  HL89	  \\
   945  &      Sm   & 15.31  &   1.29 & A &	  -9 &        8.21  &	0.31  &    3 &  HH87,D04  \\
   950  &      Sm   & 14.49  &   1.71 & A &	1098 &        8.81  &  -0.07  &    2 &  HH87	  \\
   952  &      Im   & 16.50  &   0.78 & B &	 985 &        8.12  &	0.23  &    2 &  HL89	  \\
   957  &      Sc   & 12.67  &   2.01 & S &	1695 &        8.79  &	0.02  &    1 &   M94	  \\
   958  &      Sa   & 12.13  &   3.52 & A &	-273 &        8.06  &	1.14  &    1 &  GK83,D04  \\
   963  &      Im   & 17.20  &   0.50 & A &	1866 &        7.73  &	0.01  &    3 &  HH87,D04* \\ 
   971  &      Sd   & 14.28  &   3.06 & B &	1120 &        9.26  &	0.20  &    2 &  HW89	  \\
   975  &      Scd  & 13.58  &   3.95 & B &	 933 &        9.34  &	0.33  &    1 &  HL89	  \\
   979  &      Sa   & 12.32  &   4.33 & B &	 438 &        8.40  &	1.17  &    2 &  GK83	  \\
   980  &      Scd  & 14.17  &   2.48 & A &	2342 &        8.38  &	0.67  &    1 &  HL89,D04* \\ 
   984  &      Sa   & 12.82  &   2.99 & A &	1883 &$<$     7.31  &	1.78  &    - &  GK83,D04* \\
   985  &      BCD  & 17.00  &   0.63 & S &	1638 &        7.31  &	0.61  &    3 &  HH87	  \\
   989  &      Sc   & 15.80  &   0.67 & S &	1846 &        7.50  &	0.41  &    3 &  HH87	  \\
   995  &      Sc   & 15.32  &   1.53 & A &	 928 &        8.92  &  -0.33  &    1 &  HG86	  \\
  1001  &      Im   & 16.60  &   0.73 & A &	 338 &        7.49  &	0.57  &    3 &  HH87,D04  \\
  1002  &      Sc   & 12.48  &   3.02 & B &	1450 &        8.92  &	0.47  &    1 &  HH84	  \\
  1011  &      Sdm  & 14.85  &   1.29 & S &	 874 &        8.08  &	0.43  &    1 &  HH87	  \\
  1013  &      Im   & 16.71  &   0.73 & B &	1712 &        7.51  &	0.79  &    4 &  HH87	  \\
  1017  &      Im   & 14.50  &   2.16 & B &	  32 &$<$     7.10  &	2.07  &    - &  HH87	  \\
  1021  &      Im   & 15.45  &   1.16 & B &	 868 &        7.64  &	1.03  &    4 &  HH87	  \\
  1043  &      Sb   & 10.91  &   8.12 & A &	  70 &        8.62  &	1.33  &    2 &  GK83,D04  \\
  1047  &      Sa   & 12.48  &   2.01 & A &	 724 &$<$     7.44  &	1.37  &    - &  GK83,D04* \\
  1048  &      Scd  & 15.10  &   1.71 & B &	2252 &        8.43  &	0.56  &    1 &  HL89	  \\
  1060  &      Sm   & 15.00  &   1.07 & S &	1487 &        8.10  &	0.27  &    2 &  HH87	  \\
  1091  &      Sbc  & 14.60  &   1.45 & B &	1119 &        9.30  &  -0.35  &    1 &  HG86	  \\
  1102  &      Im   & 17.70  &   0.35 & S &	   - &$<$     7.17  &	0.29  &    - &  HW89	  \\
  1106  &      Im   & 17.50  &   0.59 & A &	   - &$<$     6.64  &	1.23  &    - &  HH87	  \\
  1110  &      Sab  & 10.93  &   6.15 & A &	1954 &        8.65  &	0.95  &    1 &  HH84,D04* \\ 
  1114  &      Im   & 14.82  &   1.71 & S &	 560 &        7.28  &	1.46  &    4 &  HH87	  \\
  1118  &      Sc   & 13.31  &   1.96 & B &	 865 &        8.52  &	0.51  &    1 &  HL89	  \\
  1121  &      Im   & 16.48  &   0.71 & A &	   - &$<$     6.76  &	1.26  &    - &  T.W.	  \\
  1126  &      Sc   & 13.30  &   2.92 & A &	1687 &        7.78  &	1.33  &    1 &  HL89,D04* \\
  1128  &      Im   & 17.34  &   0.71 & B &	   - &$<$     7.24  &	1.02  &    - &  HH87	  \\
  1141  &      BCD  & 16.20  &   0.46 & B &	1040 &        7.84  &	0.07  &    3 &  HH87	  \\
  1145  &      Sb   & 11.66  &   2.92 & S &	 884 &        8.35  &	0.86  &    2 &  GK83	  \\
  1156  &      Scd  & 14.13  &   2.48 & S &	1576 &        8.82  &	0.23  &    1 &  HL89	  \\
  1158  &      Sa   & 12.09  &   3.52 & A &	1919 &$<$     7.13  &	2.07  &    - &   M94,D04* \\
  1166  &      Im   & 17.70  &   0.71 & A &	   - &$<$     7.12  &	0.90  &    - &  HH87,D04  \\
  1168  &      Im   & 17.70  &   0.43 & B &	   - &$<$     7.41  &	0.45  &    - &  HW89	  \\
  1169  &      Im   & 17.80  &   0.28 & A &	   - &$<$     6.94  &	0.33  &    - &  HW89,D04  \\
  1179  &    Im/BCD & 15.58  &   1.16 & B &	 765 &        7.73  &	0.94  &    4 &  HH87	  \\
  1189  &      Sc   & 13.70  &   1.84 & S &	 516 &        8.45  &	0.28  &    1 &  T.W.	  \\
  1190  &      Sa   & 12.22  &   4.33 & B &	 508 &$<$     7.64  &	1.93  &    - &  GK83	  \\
  1193  &      Sc   & 14.62  &   1.20 & S &	 757 &        8.43  &  -0.05  &    1 &  HG86	  \\
  1200  &      Im   & 15.10  &   1.26 & A &	-123 &$<$     6.84  &	1.65  &    - &  HH87	  \\
  1205  &      Sc   & 13.04  &   1.84 & S &	2339 &        8.76  &  -0.03  &    1 &  HH84	  \\
  1208  &      Im   & 15.20  &   0.84 & S &	1337 &        7.64  &	0.52  &    2 &  HH87	  \\
  1217  &      Sm   & 14.59  &   1.87 & A &	  38 &$<$     6.54  &	2.28  &    - &  HW89	  \\
  1249  &      Im   & 14.75  &   1.45 & S &	 468 &        7.06  &	1.55  &    4 &  HH87	  \\
  1257  &      Im   & 16.50  &   1.36 & A &	2488 &        8.41  &	0.14  &    1 &  HH87,D04* \\ 
  1266  &      Sdm  & 14.63  &   1.16 & S &	1637 &        8.24  &	0.19  &    2 &  HH87	  \\
  1273  &      Im   & 15.25  &   1.16 & B &	2015 &$<$     7.16  &	1.51  &    - &  HH87	  \\
  1287  &      Im   & 16.00  &   0.85 & A &	   - &$<$     6.80  &	1.38  &    - &  T.W.	  \\
  1290  &      Sb   & 13.09  &   2.01 & S &	2438 &        8.90  &	0.05  &    1 &  HH84	  \\
  1313  &      BCD  & 17.15  &   0.45 & A &	1254 &        7.84  &  -0.20  &    2 &  HH87	  \\
  1326  &      Sa   & 13.41  &   1.89 & A &	 497 &$<$     7.24  &	1.52  &    - &   M94	  \\
  1330  &      Sa   & 13.17  &   1.96 & S &	1777 &        7.94  &	0.85  &    2 &  HG86	  \\
  1356  &    Sm/BCD & 15.55  &   1.10 & A &	1251 &        8.34  &	0.04  &    1 &  HG86	  \\
  1358  &      Sa   & 16.00  &   0.89 & A &	   - &$<$     7.31  &	0.92  &    - &  T.W.	  \\
  1374  &    Im/BCD & 15.33  &   1.20 & A &	2555 &        8.23  &	0.23  &    2 &  HG86	  \\
  1375  &      Sc   & 12.00  &   4.76 & S &	1732 &        9.56  &  -0.05  &    1 &  HH84	  \\
  1377  &      Im   & 16.87  &   0.61 & A &	   - &$<$     6.65  &	1.25  &    - &  T.W.	  \\
  1379  &      Sc   & 12.62  &   2.85 & A &	1505 &        8.95  &	0.15  &    1 &  HH84	  \\
  1393  &      Sc   & 14.01  &   1.69 & A &	2100 &        8.43  &	0.23  &    1 &  HG86	  \\
  1401  &      Sbc  & 10.27  &   7.23 & A &	2284 &        9.36  &	0.55  &    1 &  HH84	  \\
  1403  &      Im   & 17.15  &   0.71 & A &	   - &$<$     7.02  &	1.00  &    - &  HH87	  \\
  1410  &      Sm   & 14.57  &   1.48 & A &	1629 &        8.20  &	0.42  &    1 &  HH87	  \\
  1411  &     Pec   & 15.72  &   0.70 & A &	 911 &        7.96  &	0.51  &    3 &  HH87	  \\
  1412  &      Sa   & 12.12  &   4.33 & A &	1342 &$<$     7.02  &	2.33  &    - &   M94	  \\
  1426  &      Im   & 15.64  &   0.80 & A &	1110 &$<$     6.78  &	1.34  &    - &  HH87	  \\
  1427  &    Im/BCD & 15.68  &   0.85 & A &	-132 &        7.90  &	0.27  &    3 &  HH87	  \\
  1435  &      Im   & 14.63  &   1.16 & S &	 609 &$<$     6.58  &	1.84  &    - &  T.W.	  \\
  1437  &      BCD  & 15.12  &   0.59 & S &	1148 &        8.23  &  -0.36  &    2 &  HB03	  \\
  1442  &      Sd   & 14.82  &   2.92 & S &	1735 &        8.81  &	0.37  &    1 &  HL89	  \\
  1450  &      Sc   & 13.29  &   2.60 & A &	-173 &        8.47  &	0.54  &    1 &  SS90	  \\
  1455  &      Im   & 16.80  &   0.64 & S &	1339 &        7.16  &	0.79  &    4 &  HH87	  \\
  1459  &      BCD  & 16.30  &   0.73 & S &	1774 &        7.28  &	0.77  &    4 &  HH87	  \\
  1465  &      Im   & 15.00  &   1.10 & S &	 734 &        7.54  &	0.84  &    3 &  HH87	  \\
  1468  &      Im   & 15.00  &   1.00 & S &	1233 &        8.32  &  -0.01  &    2 &  HH87	  \\
  1507  &      Sm   & 15.08  &   1.16 & S &	 910 &        8.21  &	0.22  &    2 &  HH87	  \\
  1508  &      Sc   & 12.34  &   3.60 & S &	1212 &        9.54  &  -0.26  &    1 &  HH84	  \\
  1516  &      Sbc  & 12.73  &   4.04 & S &	2330 &        8.68  &	0.80  &    1 &  HH84	  \\
  1524  &      Sd   & 13.51  &   3.20 & A &	 262 &        9.25  &	0.01  &    1 &  HL89	  \\
  1529  &      Sdm  & 14.63  &   1.16 & S &	1138 &        8.02  &	0.40  &    3 &  HH87	  \\
  1532  &      Sc   & 14.05  &   1.87 & A &	2335 &        7.93  &	0.82  &    2 &  HG86	  \\
  1540  &      Sb   & 11.32  &   5.86 & S &	1736 &        9.84  &  -0.13  &    1 &  HW89	  \\
  1552  &      Sa   & 12.58  &   4.24 & A &	 195 &$<$     7.16  &	2.18  &    - &   M94	  \\
  1554  &      Sm   & 12.30  &   2.60 & S &	2021 &        9.46  &  -0.37  &    1 &  HH84	  \\
  1555  &      Sc   & 10.51  &   8.33 & S &	1962 &        9.79  &	0.19  &    1 &  HH84	  \\
  1557  &      Scd  & 14.53  &   2.60 & S &	1759 &        8.65  &	0.44  &    1 &  HL89	  \\
  1562  &      Sc   & 11.01  &   7.23 & S &	1807 &        9.71  &	0.15  &    1 &  P79	  \\
  1566  &      Sd   & 14.80  &   1.16 & S &	 427 &        8.02  &	0.40  &    2 &  HG86	  \\
  1569  &      Scd  & 15.00  &   1.07 & A &	 799 &        7.47  &	0.90  &    2 &  HG86	  \\
  1572  &      BCD  & 16.00  &   0.93 & S &	1848 &        8.05  &	0.19  &    2 &  HH87	  \\
  1575  &      Sm   & 13.98  &   2.00 & S &	 597 &        7.94  &	0.93  &    2 &  HL89	  \\
  1581  &      Sm   & 14.55  &   1.46 & S &	2065 &        8.64  &  -0.03  &    2 &  HH87	  \\
  1585  &      Im   & 15.45  &   1.67 & A &	 666 &        8.79  &  -0.07  &    1 &  HH87	  \\
  1588  &      Scd  & 12.81  &   2.60 & A &	1288 &        8.40  &	0.68  &    1 &  SS90	  \\
  1596  &      Im   & 17.24  &   0.35 & S &	1286 &        7.34  &	0.12  &    4 &  HH87	  \\
  1597  &      Sc   & 15.20  &   0.93 & S &	 841 &        7.54  &	0.63  &    1 &  T.W.	  \\
  1605  &      Sd   & 17.00  &   1.00 & A &	1077 &        7.74  &	0.56  &    2 &  HL89	  \\
  1615  &      Sb   & 10.98  &   6.00 & A &	 484 &        8.93  &	0.80  &    1 &  W86	  \\
  1624  &      Sc   & 13.89  &   2.48 & S &	1151 &        8.34  &	0.64  &    1 &  HW89	  \\
  1644  &      Sm   & 17.50  &   0.98 & A &	 756 &        8.15  &	0.14  &    1 &  HH87	  \\
  1654  &      Im   & 15.96  &   0.85 & A &	2051 &        8.13  &	0.04  &    2 &  HH87	  \\
  1673  &      Sc   & 12.08  &   2.92 & A &	2277 &        8.69  &	0.43  &    3 &  HS82	  \\
  1675  &     Pec   & 14.47  &   1.26 & S &	1795 &        7.45  &	1.35  &    3 &  HH87	  \\
  1676  &      Sc   & 11.70  &   5.10 & A &	2255 &        8.99  &	0.58  &    3 &  HS82	  \\
  1678  &      Sd   & 13.70  &   2.16 & S &	1073 &        9.00  &  -0.06  &    2 &  HL89	  \\
  1685  &      Sd   & 15.18  &   2.16 & S &	1443 &        8.61  &	0.32  &    1 &  HL89	  \\
  1686  &      Sm   & 13.95  &   2.79 & A &	1122 &        8.35  &	0.79  &    1 &  HH87	  \\
  1690  &      Sab  & 10.25  &  10.73 & A &	-216 &        8.93  &	1.07  &    1 &  HR89	  \\
  1696  &      Sc   & 11.81  &   4.58 & A &	 342 &        8.94  &	0.54  &    1 &  HH84	  \\
  1699  &      Sm   & 14.11  &   1.55 & S &	1635 &        8.62  &	0.04  &    2 &  HH87	  \\
  1725  &    Sm/BCD & 14.51  &   1.55 & S &	1068 &        8.11  &	0.55  &    2 &  HH87	  \\
  1726  &      Sdm  & 14.54  &   1.29 & S &	  61 &        8.52  &	0.00  &    2 &  HH87	  \\
  1727  &      Sab  & 10.56  &   6.29 & A &	1520 &        8.79  &	0.83  &    1 &  HS81	  \\
  1728  &      Im   & 16.63  &   0.50 & E &	-117 &        7.35  &	0.38  &    5 &  HH87	  \\
  1730  &      Sc   & 12.61  &   2.16 & S &	1032 &        7.83  &	1.03  &    3 &  HH84	  \\
  1744  &      BCD  & 17.50  &   0.51 & E &	1150 &        7.16  &	0.60  &    2 &  HW89	  \\
  1750  &      BCD  & 16.50  &   0.31 & S &	-117 &        7.35  &  -0.01  &    4 &  HH87	  \\
  1753  &      Im   & 16.81  &   0.71 & A &	 737 &        7.70  &	0.32  &    4 &  HH87	  \\
  1757  &      Sa   & 13.60  &   1.87 & A &	1783 &        7.38  &	1.38  &    5 &  HG86	  \\
  1758  &      Sc   & 14.99  &   1.71 & S &	1788 &        8.28  &	0.39  &    1 &  HL89	  \\
  1760  &      Sa   & 12.54  &   4.33 & S &	 792 &        8.20  &	1.16  &    1 &  HH84	  \\
  1771  &      Im   & 17.97  &   0.31 & E &	   - &$<$     7.02  &	0.34  &    - &  HW89	  \\
  1780	&      Sb   & 13.70  &   1.87 & S &	2424 &        8.41  &	0.48  &    1 &  HL89	  \\
  1784  &      Im   & 15.84  &   0.79 & E &	  57 &        7.34  &	0.78  &    4 &  HH87	  \\
  1789  &      Im   & 15.07  &   1.10 & S &	1619 &        7.86  &	0.53  &    1 &  HH87	  \\
  1791  &    Sm/BCD & 14.67  &   1.29 & S &	2079 &        8.63  &  -0.12  &    2 &  HH87	  \\
  1804  &    Im/BCD & 15.63  &   0.75 & E &	1898 &        7.23  &	0.84  &    5 &  HH87	  \\
  1811  &      Sc   & 12.92  &   2.16 & E &	 632 &        8.63  &	0.23  &    1 &  HH84	  \\
  1813  &      Sa   & 11.51  &   4.76 & E &	1834 &$<$     7.19  &	2.23  &    - &   M94	  \\
  1816  &      Im   & 16.20  &   1.16 & E &	1002 &        8.31  &	0.12  &    2 &  HH87	  \\
  1822  &      Im   & 15.60  &   0.63 & S &	1012 &        7.64  &	0.29  &    2 &  HH87	  \\
  1859  &      Sa   & 12.52  &   5.10 & E &	1645 &        7.81  &	1.66  &    2 &   M94	  \\
  1868  &      Scd  & 13.75  &   3.95 & E &	2255 &        8.53  &	0.89  &    1 &  HG86	  \\
  1885  &      Im   & 16.41  &   1.16 & E &	   - &$<$     6.52  &	1.90  &    - &  T.W.	  \\
  1918  &      Im   & 15.80  &   1.03 & S &	 980 &        8.16  &	0.17  &    2 &  HH87	  \\
  1923  &      Sbc  & 13.14  &   2.31 & S &	 742 &        8.62  &	0.45  &    1 &  HL89	  \\
  1929  &      Scd  & 13.77  &   2.48 & E &	 291 &        8.70  &	0.35  &    1 &  SS90	  \\
  1931  &      Im   & 15.20  &   1.26 & E &	1100 &        8.40  &	0.10  &    3 &  HH87	  \\
  1932  &      Sc   & 13.19  &   2.92 & E &	 116 &        8.66  &	0.45  &    1 &   M94	  \\
  1933  &      Sm   & 15.80  &   0.71 & S &	2409 &        8.02  &	0.00  &    2 &  HL89	  \\
  1943  &      Sb   & 12.19  &   3.20 & E &	1048 &        9.03  &	0.25  &    1 &  HS81	  \\
  1952  &      Im   & 16.00  &   0.71 & E &	1308 &        8.21  &  -0.19  &    2 &  HH87	  \\
  1955  &    S/BCD  & 14.32  &   1.36 & E &	2012 &        7.72  &	0.83  &    3 &  HH87	  \\
  1960  &    Im/BCD & 17.00  &   0.46 & E &	   - &$<$     7.09  &	0.58  &    - &  HH87	  \\
  1965  &      Im   & 16.50  &   0.50 & S &	 954 &        7.30  &	0.44  &    4 &  HH87	  \\
  1970  &      Im   & 15.80  &   0.71 & E &	1324 &        6.95  &	1.07  &    3 &  T.W.	  \\
  1972  &      Sc   & 12.03  &   2.60 & E &	1422 &        8.75  &	0.27  &    1 &  HH84	  \\
  1987  &      Sc   & 11.14  &   4.99 & E &	1039 &        9.85  &  -0.29  &    1 &  HS81	  \\
  1992  &      Im   & 15.50  &   0.81 & E &	1003 &        8.35  &  -0.22  &    1 &  HH87	  \\
  1999  &      Sa   & 13.08  &   1.99 & E &	 510 &$<$     7.19  &	1.61  &    - &   M94	  \\
  2007  &    Im/BCD & 15.20  &   0.78 & E &	1857 &        7.37  &	0.74  &    3 &  HH87	  \\
  2015  &      BCD  & 16.20  &   0.51 & E &	2545 &$<$     7.09  &	0.67  &    - &  HH87	  \\
  2023  &      Sc   & 13.86  &   2.01 & E &	 958 &        8.85  &  -0.05  &    1 &  HG86	  \\
  2033  &      BCD  & 14.65  &   0.73 & E &	1486 &        7.45  &	0.61  &    3 &  HH87	  \\
  2034  &      Im   & 15.82  &   0.78 & E &	1500 &        7.83  &	0.27  &    3 &  HH87	  \\
  2037  &    Im/BCD & 15.92  &   0.88 & E &	1142 &        7.39  &	0.81  &    3 &  HH87	  \\
  2058  &      Sc   & 11.55  &   5.86 & E &	1620 &        8.79  &	0.90  &    1 &  HH84	  \\
  2070  &      Sa   & 11.53  &   5.67 & E &	1008 &        9.54  &	0.01  &    1 &  HH84	  \\
  2089  &      BCD  & 17.50  &   0.39 & E &	   - &$<$     7.09  &	0.46  &    - &  HH87	  \\
  2094  &      Im   & 17.80  &   0.37 & E &	   - &$<$     6.54  &	0.95  &    - &  HW89	  \\

   \noalign{\smallskip}
   \hline
\multicolumn{9}{l}{Column 1: VCC designation.}\\  
\multicolumn{9}{l}{Column 2: Morphological type from Binggeli et al. (1985; 1993);}\\
\multicolumn{9}{l}{Column 3: Apparent photographic magnitude from Binggeli et al. (1985);}\\
\multicolumn{9}{l}{Column 4: Optical major diameter, in arcmin;}\\
\multicolumn{9}{l}{Column 5: Subcloud membership as in Gavazzi et al. (1999b);}\\ 
\multicolumn{9}{l}{Column 6: Recessional velocity, in \kms;}\\
\multicolumn{9}{l}{Column 7: \HI\ mass or mass limit in solar units:}\\
\multicolumn{9}{l}{Column 8: \HI\ deficiency parameter as defined in Haynes \& Giovanelli (1984);}\\
\multicolumn{9}{l}{Column 9: Quality flag (see last Column of Table 1).}\\
\multicolumn{9}{l}{Column 10: Reference to the \HI\ measurement;}\\

   \setcounter{table}{2}
   \label{sample_dat}
   \end{longtable}
}

\section{Appendix A: Notes on individual galaxies}

\normalsize

{\sf  VCC 48:} The possible line emission of this object, which has an optical redshift of 
$-$52$\pm$60\kms, is masked by Galactic \HI\ emission.
The object remains undetected in spite of the noise reduction from 1.1 mJy,
as reported by Hoffman et al. (1987), to 0.64 mJy rms in the present work.
\\
{\sf VCC 222:} 
We did not detect this galaxy, with an rms noise level of 0.60 mJy.
Its optical redshift, 2298$\pm$122 \kms, is not well determined.
An Effelsberg \HI\ detection was reported by Huchtmeier (1982) at 2410 \kms, with
$W_{50}$=273 \kms\ and $I_{HI}$=4.4$\pm$1.5 \Jykms, with an average line signal of 16 mJy. 
Although Magri (1994) reported a tentative detection at 2596 \kms, its low signal-to-noise 
ratio makes it appear spurious. Two published (Krumm \& Salpeter 1979; Mirabel \& Wilson 1984) 
estimated upper limits to its line flux are 2.4 and 3.4 \Jykms, respectively. 
We conclude that the line signal reported by Huchtmeier is spurious and due to RFI.
\\
{\sf VCC 227, 256:} Marginal detections.
\\
{\sf VCC 323:}
The object remains undetected in spite of the noise reduction from 6.0 mJy,
as reported by Huchtmeier \& Richter (1986), to 0.53 mJy rms in the present work.
\\
{\sf VCC 341:} We did not detect this galaxy, with an rms noise level of 0.76 mJy.
Its optical redshift is 1827$\pm$60 \kms.
The Effelsberg \HI\ detection reported by Huchtmeier (1982) at 1846 \kms, with
$W_{50}$=465 \kms\ and $I_{HI}$=3.0$\pm$1.5 \Jykms, i.e. an average line strength of
6.5 mJy, appears spurious given its low signal-to-noise ratio.
\\
{\sf VCC 358:} Bad baselines due to the proximity of the 19 Jy continuum source NGC 4261
resulted in the very high rms noise level of 28 mJy, about 30 times
that of similar observations made for this project. Our previous \nan\ observations
(van Driel et al. 2000) also had a high rms noise level of 11 mJy and a limit
of 6 \Jykms\ was reported by Huchtmeier \& Richter (1986) from Effelsberg data.
An Arecibo detection at 2633 \kms\ with \IHI=9.0 \Jykms\ was reported by Magri (1994).
\\
{\sf VCC 362:} Given that its optical redshift is 156$\pm$44 \kms,
our \HI\ detection ($V_{HI}$=2350 \kms, $W_{50}$=92 \kms\ and $I_{HI}$=0.40 \Jykms)
must be due to confusion.
At  \am{4}{9} lies VCC 367, a 17.4 mag object without published optical redshift, 
for which Hoffman et al. (1987) measured 
$V_{HI}$=2362 \kms, $W_{50}$=98 \kms\ and $I_{HI}$=0.59 \Jykms\ at Arecibo.
VCC 362 remains undetected in spite of the noise reduction from 3.5 mJy,
as reported by Magri (1994), to 0.44 mJy rms in the present work.
\\
{\sf VCC 524:} Besides a detection of the target galaxy at 1035 \kms, our spectrum
shows a detection near 5800 \kms with $I_{HI}$=0.45 \Jykms, 
which is due to confusion by VCC 526 (= NGC 4307A), a 15.7 mag Sc spiral \am{3}{2} away, 
with $V_{opt}$=5989$\pm$46 \kms, for which Magri (1994) measured $V_{HI}$=5836 \kms, 
$W_{50}$=76 \kms\ and $I_{HI}$=2.6 \Jykms\ at Arecibo.
 \\
{\sf VCC 528:} is a background galaxy, at 7046 \Jykms.
\\
{\sf VCC 666:} The possible detection at -250 \kms is doubtful.
The object remains undetected in spite of the noise reduction from 1.3 mJy,
as reported by Hoffman et al. (1987), to 0.51 mJy rms in the present work.
\\
{\sf VCC 802:}
The object remains undetected in spite of the noise reduction from 1.5 mJy,
as reported by Hoffman et al. (1987), to 0.58 mJy rms in the present work.
\\
{\sf VCC 1086:} its optical redshift is 294$\pm$39 \kms.
Our \HI\ parameters ($V_{HI}$=328$\pm$8 \kms, $W_{50}$=240 \kms\ and $I_{HI}$=0.93$\pm$0.09 \Jykms)
are comparable to those ($V_{HI}$=363 \kms, $W_{50}$=224 \kms\ and $I_{HI}$=0.96 \Jykms)
measured by Hoffman et al. (1989a) at Arecibo. Profile contaminated by galactic absorption.
\\
{\sf VCC 1189:} its optical redshift is 544$\pm$43 \kms.
Our \HI\ profile has $V_{HI}$=516$\pm$1 \kms, $W_{50}$=112 \kms\ and $I_{HI}$=4.2$\pm$0.1 \Jykms.
The profile parameters measured by, respectively, Huchtmeier \& Richter (1986) at Effelsberg, 
Hoffman et al. (1989a) at Arecibo and Schneider et al. (1992) with the NRAO 300ft are 
$V_{HI}$=530, 537 and 543 \kms, $W_{50}$=98, 110 and 121 \kms\ and 
$I_{HI}$=5.0, 3.6 and 9.7 \Jykms. Although the line fluxes vary considerably, they 
do not correlate with the size of the telescope beam, nor are there any candidates for 
confusion in the vicinity. The large line flux measured by Schneider et al. appears to be
spurious.
\\
{\sf VCC 1196:} We did not detect this galaxy, with an rms noise level of 0.57 mJy.
Its optical redshift is 908$\pm$36 \kms.
An Effelsberg \HI\ detection was reported by Huchtmeier \& Richter (1986) at 2422 \kms, with
$W_{50}$=446 \kms\ and $I_{HI}$=6.5$\pm$1.5 \Jykms, i.e. an average line strength of 15 mJy,
which we assume is spurious and due to RFI. 
\\
{\sf VCC 1287:}
The object remains undetected in spite of the noise reduction from 1.4 mJy,
as reported by Hoffman et al. (1987), to 0.77 mJy rms in the present work.
\\
{\sf VCC 1377:}
The object remains undetected in spite of the noise reduction from 1.1 mJy,
as reported by Hoffman et al. (1987), to 0.51 mJy rms in the present work.
\\
{\sf VCC 1435:}
The object remains undetected in spite of the noise reduction from 0.9 mJy,
as reported by Hoffman et al. (1987), to 0.44 mJy rms in the present work.
\\
{\sf VCC 1448:} We did not detect this galaxy, which does not have a published optical
redshift, with an rms noise level of 0.92 mJy.
An Effelsberg \HI\ detection was reported by Huchtmeier \& Richter (1986) at 2583 \kms, with
$W_{50}$=66 \kms\ and $I_{HI}$=1.5$\pm$0.26 \Jykms, i.e. with an average line signal of 23 mJy,
which we assume is spurious and due to RFI. 
\\
{\sf VCC 1885:}
The object remains undetected in spite of the noise reduction from 0.7 mJy,
as reported by Hoffman et al. (1987), to 0.38 mJy rms in the present work.
\\
{\sf VCC 1970:}
Hoffman et al. (1987) reported undetection with 1.0 mJy rms. We detected it
at a noise level of 0.37 mJy rms.


\end{document}